\documentclass[10pt,journal]{IEEEtran}

\usepackage{amsmath}
\usepackage{amssymb}
\usepackage{bm}
\usepackage{graphicx}
\usepackage{psfrag}
\usepackage{subfigure}
\usepackage{caption}
\usepackage{color}
\usepackage{comment}
\usepackage[dvipsnames]{xcolor}
\usepackage{paralist}
\usepackage{enumitem}
\usepackage{relsize}
\usepackage{hyperref}
\usepackage{tabularx}

\DeclareMathAlphabet{\pazocal}{OMS}{zplm}{m}{n}

\newcommand{\dps}{\displaystyle}
\newcommand\norm[1]{\left\lVert#1\right\rVert}

\newcommand{\fone}{\hat{f}^{\rm t1}}
\newcommand{\fpone}{\hat{f}^{\rm t1,p}}
\newcommand{\fsone}{\hat{f}^{\rm t1,s}}
\newcommand{\ftwo}{\hat{f}^{\rm t2}}

\newcommand{\np}{n^{\rm p}}

\newcommand{\nx}{n^{\bm x}}

\newcommand{\xm}{\bm{x}^{\rm m}}
\newcommand{\xe}{\bm{x}^{\rm e}}

\newcommand{\Nu}{n^{\bm u}}

\newcommand{\Kid}{K^{\rm id}}
\newcommand{\Ktune}{K^{\rm tune}}
\newcommand{\Km}{K^{\rm m}}
\newcommand{\Kc}{K^{\rm c}}
\newcommand{\tx}{\bm{\theta^x}}
\newcommand{\fx}{\hat{f}^{\bm{x}}}
\newcommand{\tu}{\bm{\theta^u}}
\newcommand{\fu}{\hat{f}^{\bm{u}}}
\newcommand{\txcon}{\bm{\theta}^{{\bm x}, {\rm con}}}
\newcommand{\txant}{\bm{\theta}^{{\bm x}, {\rm ant}}}
\newcommand{\tucon}{\bm{\theta}^{{\bm u}, {\rm con}}}
\newcommand{\tuant}{\bm{\theta}^{{\bm u}, {\rm ant}}}
\newcommand{\delid}{\delta^{\rm id}}
\newcommand{\pid}{\pi^{\rm id}_s}
\newcommand{\ptune}{\pi^{\rm tune}_s}
\newcommand{\Lid}{L^{\rm id}}
\newcommand{\Ltune}{L^{\rm tune}}

\newcommand{\xcor}[1]{\bm{x}_{#1}}
\newcommand{\vcor}[1]{\bm{\nu}_{#1}}

\newcommand{\pz}[1]{\pi_{\mathsmaller{0}}\left(#1\right)}
\newcommand{\pone}[1]{\pi_{\mathsmaller{1}}\left(#1\right)}
\newcommand{\ptwo}[1]{\pi_{\mathsmaller{2}}\left(#1\right)}
\newcommand{\pdelta}[1]{\pi_{\mathsmaller{\delta}}\left(#1\right)}

\newcommand{\kz}{k_{\mathsmaller{0}}}

\newcommand{\deltune}{\delta^{\rm tune}}
\newcommand{\ns}{n^{\rm s}}

\newcommand{\ueq}{{\hat{\pazocal{U}}}^{\text{\tiny eq}}}
\newcommand{\uneq}{{\hat{\pazocal{U}}}^{\text{\tiny neq}}}

\newcommand{\event}[2]{e^{#1}_{#2}}
\newcommand{\low}{{\rm low}}
\newcommand{\high}{\rm high}
\newcommand{\short}{{\rm short}}
\newcommand{\lng}{{\rm long}}

\newtheorem{remark}{Remark}

\makeatletter
\newcommand{\subalign}[1]{%
  \vcenter{%
    \Let@ \restore@math@cr \default@tag
    \baselineskip\fontdimen10 \scriptfont\tw@
    \advance\baselineskip\fontdimen12 \scriptfont\tw@
    \lineskip\thr@@\fontdimen8 \scriptfont\thr@@
    \lineskiplimit\lineskip
    \ialign{\hfil$\m@th\scriptstyle##$&$\m@th\scriptstyle{}##$\crcr
      #1\crcr
    }%
  }
}
\makeatother

\ifCLASSOPTIONcompsoc
\usepackage[nocompress]{cite}
\else
 \usepackage{cite}
\fi

\ifCLASSINFOpdf
\else
\fi

\makeatletter
\newcommand{\bigger}{\bBigg@{4}}
\makeatother

\allowdisplaybreaks
\begin{document}

\setlength{\abovedisplayskip}{3pt}
\setlength{\belowdisplayskip}{3pt}

\setlength{\textfloatsep}{4pt plus 1.0pt minus 2.0pt}

\setlength{\medmuskip}{0mu}
\setlength{\thickmuskip}{0mu}
\setlength{\thinmuskip}{0mu}


\newcommand{\papertitle}{\huge{Integrated intelligent and predictive control: A multi-agent adaptive 
type-2 fuzzy control architecture}}

\title{\papertitle}

\author{Anahita~Jamshidnejad*,
		Emilio Frazzoli,         
                Mohammad~J.~Mahjoob, and
                Bart~De~Schutter\vspace{-5.5ex}
                \thanks{\IEEEcompsocthanksitem A.\ Jamshidnejad
                   and E.\ Frazzoli are with the Department of
                   Mechanical and Process Engineering, Institute for
                   Dynamic Systems and Control, ETH
                   Z\"urich, 
                   Switzerland.  M.J.~Mahjoob is with the
                   Faculty of Mechanical Engineering, University of
                   Tehran, 
                   Iran.  B.\ De~Schutter is with the 
                   Delft Center for Systems and Control, 
                   Delft University of Technology, 
                   The Netherlands.}  \thanks{* Corresponding author:
                   \href{mailto:ana.jamshidnejad@gmail.com}{ana.jamshidnejad@gmail.com}.}
}

\markboth{}{}  


\maketitle

\begin{abstract}


We propose a novel two-layer multi-agent architecture aimed at efficient 
real-time control of large-scale and complex-dynamics systems.  
The proposed architecture integrates intelligent control approaches (which have a low computation time and 
fit real-time applications) with model-predictive control (which takes care of the 
optimality requirements of control). 
The bottom control layer (intelligent-control module) includes several distributed 
intelligent-control agents, the design parameters of which are tuned by the top layer 
(model-predictive control module). 
The model-predictive control module fulfills two significant roles:  
looking ahead to the effects of the control decisions, and coordinating the intelligent-control agents of the lower control layer. 
The resulting multi-agent control system has a very \emph{low computation time}, and provides 
\emph{adaptivity}, \emph{control coordination}, and aims at \emph{excellent performance}. 
Additionally, we give a general treatment of type-2 fuzzy membership functions, and introduce two 
categories for them: probabilistic-fuzzy (which is a novel concept introduced in this paper) and 
fuzzy-fuzzy (which is a new treatment of the existing type-2 fuzzy membership functions). 
The performance of the proposed modeling and control approaches are assessed via a case study 
involving a simple urban traffic network:  
the results show that the novel concept of probabilistic-fuzzy membership function outperforms the type-1 
and type-2 membership functions that have already been introduced in the literature. 
Furthermore, the proposed two-layer integrated multi-agent control architecture 
significantly outperforms a multi-agent decentralized fuzzy control system 
(without coordination among the agents), 
while requiring a comparable computation time.%
  
\end{abstract}


\IEEEpeerreviewmaketitle

\section{Motivation}
\label{sec:introduction}

\IEEEPARstart{M}{ulti-agent} control systems have been developed to tackle control problems of systems with 
large spatial scales and/or complex dynamics. 
These control problems should be addressed by control systems that provide 
\emph{flexibility} for responding to the various control requirements for the different 
spatial scales and dynamical elements.  
Such a flexibility is best provided by multi-agent systems \cite{Sycara:1998, Olivera:1999}.%

Multi-agent control systems have been used for various applications, such as transportation and power networks 
\cite{Negenborn:2008, Negenborn:2010}, robotic teams \cite{Tambe:1998}, 
traffic networks \cite{Jamshidnejad:2015}, and chilled-water systems \cite{Maturana:2005}. 
Based on these works, the main open challenges with multi-agent 
control systems involve
\begin{itemize}
\item 
Optimality/suboptimality guarantees as a consequence of the collective behavior of agents.
\item
Real-time control computations (i.e., computations that do not take more time than one control sampling cycle). 
\item
Adaptivity with respect to external disturbances and unexpected/unpredicted changes in the system's dynamics.
\item
Effective coordination of agents, both in terms of the influence of the current 
control action on the current performance, as well as on the near-future performance.  
\end{itemize}%

To address these challenges, we propose a novel two-layer control
approach that integrates intelligent and predictive control methods
within a multi-agent architecture.  Intelligent control approaches
usually involve a very low computation time, making them suitable for
real-time applications. 
They can cope with high levels of nonlinearity in the dynamics of the
controlled system and can be designed in an adaptive or self-organized
way \cite{Harris:1994}.  Model-based predictive control approaches
involve optimization-based methods that minimize a predefined cost
function within a finite prediction window 
\cite{Maciejowski:2002}. The control decision is made based on
predicted values of the state of the controlled system, and hence
involves the potential future effects of the control actions on the
controlled system. %
The integrated two-layer control system proposed in this paper
possesses the following 
characteristics:
\begin{compactitem}
\item
Potential for implementation to systems with large spatial scales or several complex elements
\item
Very low computation time 
\item
High levels of adaptivity 
\item
Effective coordination among various control agents  
\item
Taking into account the future dynamics (MPC module) 
and adaptively improving performance by considering past dynamics (intelligent-control module).
\end{compactitem}
\smallskip

\noindent
\textit{Contributions and organization of the paper}
\\
%
The main contributions of the paper include:
\begin{itemize}[leftmargin=*]
\item 
We present an extensive treatment of type-2 fuzzy sets and membership functions 
that is more general than the ones that can be found in literature.   
Two forms of type-2 membership functions, called probabilistic-fuzzy (a novel concept introduced in 
this paper) and fuzzy-fuzzy (a new treatment of the existing type-2 membership functions), 
are introduced.
\item A novel two-layer integrated control architecture is proposed.
   Multi-agent control, model-based intelligent control, and
   model-predictive control are combined to obtain a control system
   with a low computation time, providing adaptivity and coordination
   among various agents.
\item We introduce a general formulation of type-2 fuzzy rules for
   modeling dynamics influenced by both delayed and current
   inputs: Intelligent-control agents in the proposed bi-level
   architecture use this formulation both for decision making and for
   estimation of the past missing states, while the MPC module uses
   the same formulation in its prediction model.
\item The proposed integrated modeling and control framework is
   implemented to and evaluated for an urban traffic network.
\end{itemize}
\smallskip
\noindent
The paper is organized as follows:
Section~\ref{sec:background} provides a brief background discussion.
Section~\ref{sec:Novel_formulation} discusses the novel formulation
for developing fuzzy rules, as well as two forms of type-2 membership
functions.  Section~\ref{sec:two_layer_controller} explains the
proposed two-layer integrated control architecture, where
Section~\ref{sec:first_control_layer} details the bottom layer of
control, including the fuzzy model and the fuzzy-control agents of the
subsystems, 
and Section~\ref{sec:MPC_controller} explains the top layer of control
(consisting of the MPC module).
Section~\ref{sec:case_study_and_results} presents the results of a
case study for an urban traffic network.
Table~\ref{table:mathematical_notations}
gives the frequently-used mathematical notations.%

\setlength{\tabcolsep}{5pt}
\begin{table}
\caption{Frequently-used mathematical notations.}
\label{table:mathematical_notations}
\begin{tabularx}{\linewidth}{l|X}
\hline
$\bm{u}_{i,s}$ & $i^{\rm th}$ input vector of subsystem $s$\\
$\bm{u}_s$ & vector of input vectors $\bm{u}_{i,s}$ for all $i$ \\
$\bm{u}_{\ell,i,s}$ & $\ell^{\rm th}$ control input that affects $i^{\rm th}$ 
state variable of subsystem $s$ \\
$\xm_{i,s}$ & measured $i^{\rm th}$ state variable of subsystem $s$ \\
$\xm_s$ & vector of measured state variables $\xm_{i,s}$ for all $i$ \\
$\xm_{\ell,i,s}$ & measured $\ell^{\rm th}$ state variable of subsystem $s$ affected by $\bm{u}_{i,s}$ \\
$\xe_{i,s}$ & estimated $i^{\rm th}$ state variable of subsystem $s$ \\
$\xe_s$ & vector of estimated state variables $\xe_{i,s}$ for all $i$ \\
$\xe_{\ell,i,s}$ & estimated $\ell^{\rm th}$ state variable of subsystem $s$ affected by $\bm{u}_{i,s}$ \\
$\bm{\nu}_{i,s}$ & external disturbances for subsystem $s$ affecting $i^{\rm th}$ state variable \\
$\bm{\nu}_s$ & vector of external disturbances $\bm{\nu}_{i,s}$ for all $i$ \\
$\bm{\nu}_{\ell,i,s}$ & external disturbances for subsystem $s$ affecting $\xm_{\ell,i,s}$ \\
$\Kc$ & set of all control time steps \\
$\Km$ & set of time steps with a reliable measurement of state variable \\
$\Kid$ & set of identification time steps for subsystem $s$ \\
$\Ktune$ & set of tuning time steps for subsystem $s$ \\
$\kappa(k)$ & average step cost value at time step $k$ \\
$\hat{J}(\cdot)$ & step cost function \\
$\phi_s$ & optimal future cumulative local cost for subsystem $s$ by MPC \\
\hline
\end{tabularx}
\end{table}


\section{Background}
\label{sec:background}

Now we provide a brief discussion on 
fuzzy logic and uncertainties.%

%

\subsection{Fuzzy logic}
\label{sec:intro:FL}

Among different intelligent control approaches, the main focus of this paper is on fuzzy logic-based  
control, because fuzzy logic converts user-supplied rules formulated in vague human language
into mathematical equivalents, can handle problems with imprecise or incomplete information, 
can model nonlinear and complex functions, while the corresponding rules are easily adaptable and maintainable over time.%


In classical logic, a realized value of the variable $x$ either ``belongs to" a crisp set 
(membership degree of 1) or ``does not belong to" it (membership degree of 0), 
while in fuzzy logic, any realized value of $x$ can belong to a fuzzy set with a certain  
membership degree within $[0,1]$, determined by a membership function. 
Each value $x^*$ adopts a single crisp membership degree $\fone\left(x^*\right)$ via a type-1 
fuzzy membership function.

\begin{figure}
\begin{center}
\psfrag{x*}[][][.7]{$x^*$}
\psfrag{w}[][][.7]{$\mu_2$}
\psfrag{mu}[][][.7]{$\mu_1$}
\psfrag{x}[][][.7]{$x$}
\psfrag{mu11}[][][.7]{$\mu_{1,1}$}
\psfrag{mu12}[][][.7]{$\mu_{1,2}$}
\psfrag{mu13}[][][.7]{$\mu_{1,3}$}
\psfrag{mu2mu11}[][][.7]{$\mu_{2,1}$}
\psfrag{mu2mu12}[][][.7]{$\mu_{2,2}$}
\psfrag{mu2mu13}[][][.7]{$\mu_{2,3}$}
\includegraphics [width=.7\linewidth] {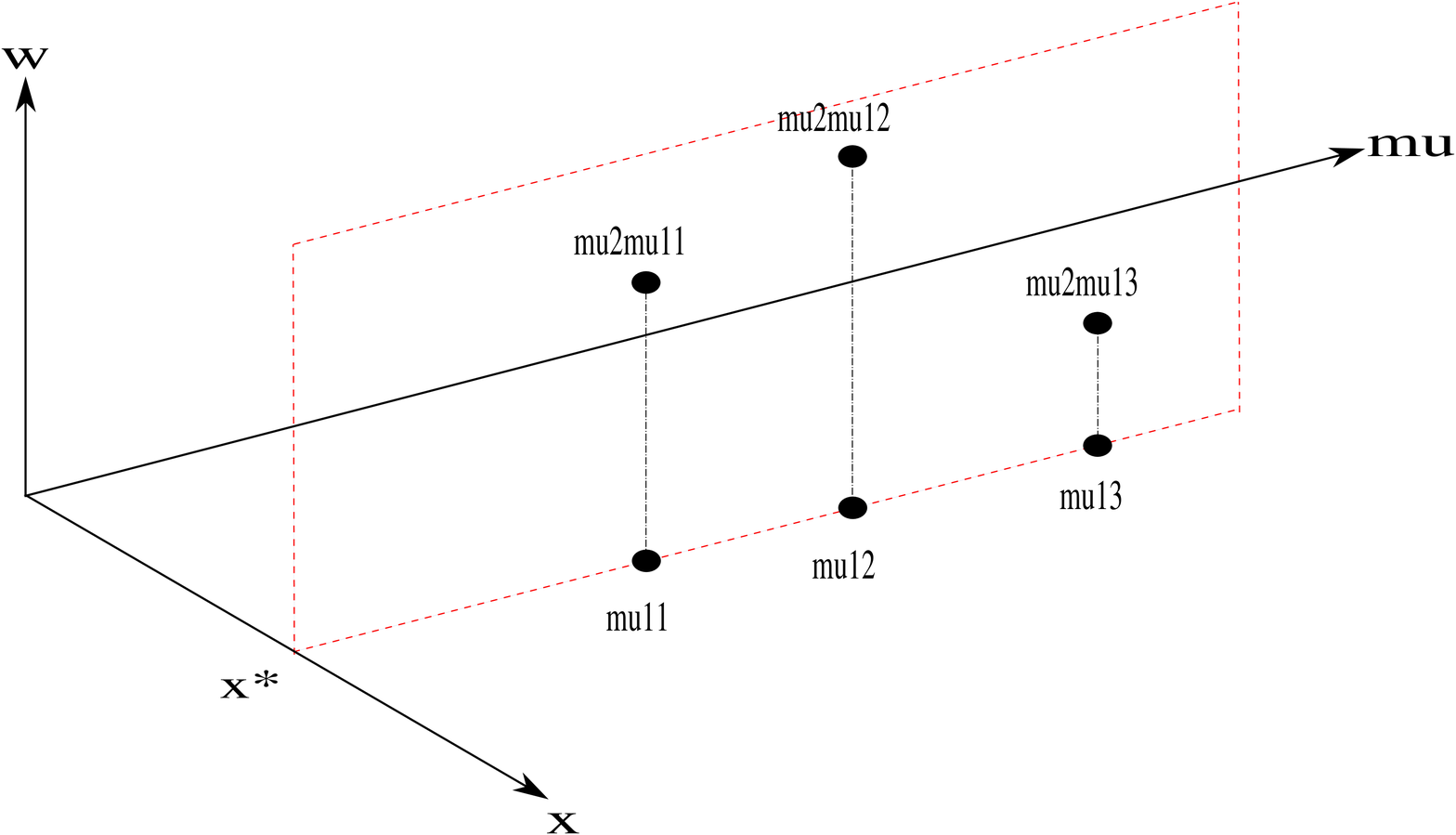}
\caption{Type-2 membership function (discrete-time domain).}
\label{fig:Primary_Secondary_MV}
\vspace*{2ex}
%
\psfrag{sfu}[][][.7]{}
\psfrag{pfu}[][][.7]{}
\psfrag{pf1}[][][.7]{$\fone_{{\rm p},1}$}
\psfrag{pfn}[][][.7]{$\fone_{{\rm p},n\rightarrow \infty}$}
\psfrag{-3}[][][1]{}
\psfrag{-2}[][][1]{}
\psfrag{-1}[][][1]{}
\psfrag{0}[][][1]{}
\psfrag{1}[][][1]{}
\psfrag{2}[][][1]{}
\psfrag{3}[][][1]{}
\psfrag{4}[][][1]{}
\psfrag{5}[][][1]{}
\psfrag{6}[][][1]{}
\psfrag{7}[][][1]{}
\psfrag{8}[][][1]{}
\psfrag{9}[][][1]{}
\psfrag{mu10}[][][.7]{$\mu_{1,0}$}
\psfrag{mu1n}[][][.7]{$\mu_{1,n}$}
\psfrag{zero}[][][.6]{0}
\psfrag{one}[][][.6]{1}
\psfrag{x}[][][.8]{$x$}
\psfrag{x*}[][][.8]{$x^*$}
\psfrag{mu1}[][][.8]{\hspace{1ex}$\mu_1$}
\psfrag{mu2}[][][.8]{$\mu_2$}
\psfrag{ft2}[][][.8]{$
\begin{array}{l}
\fsone  \left(x^*,\mu_1^*\right),\\ 
\mu_1^*\in\left[\mu_{1,0},\mu_{1,n}\right]
\end{array}
$}
\psfrag{t2}[][][.7]{Type-2 membership function $\ftwo$}
\includegraphics [width=.9\linewidth] {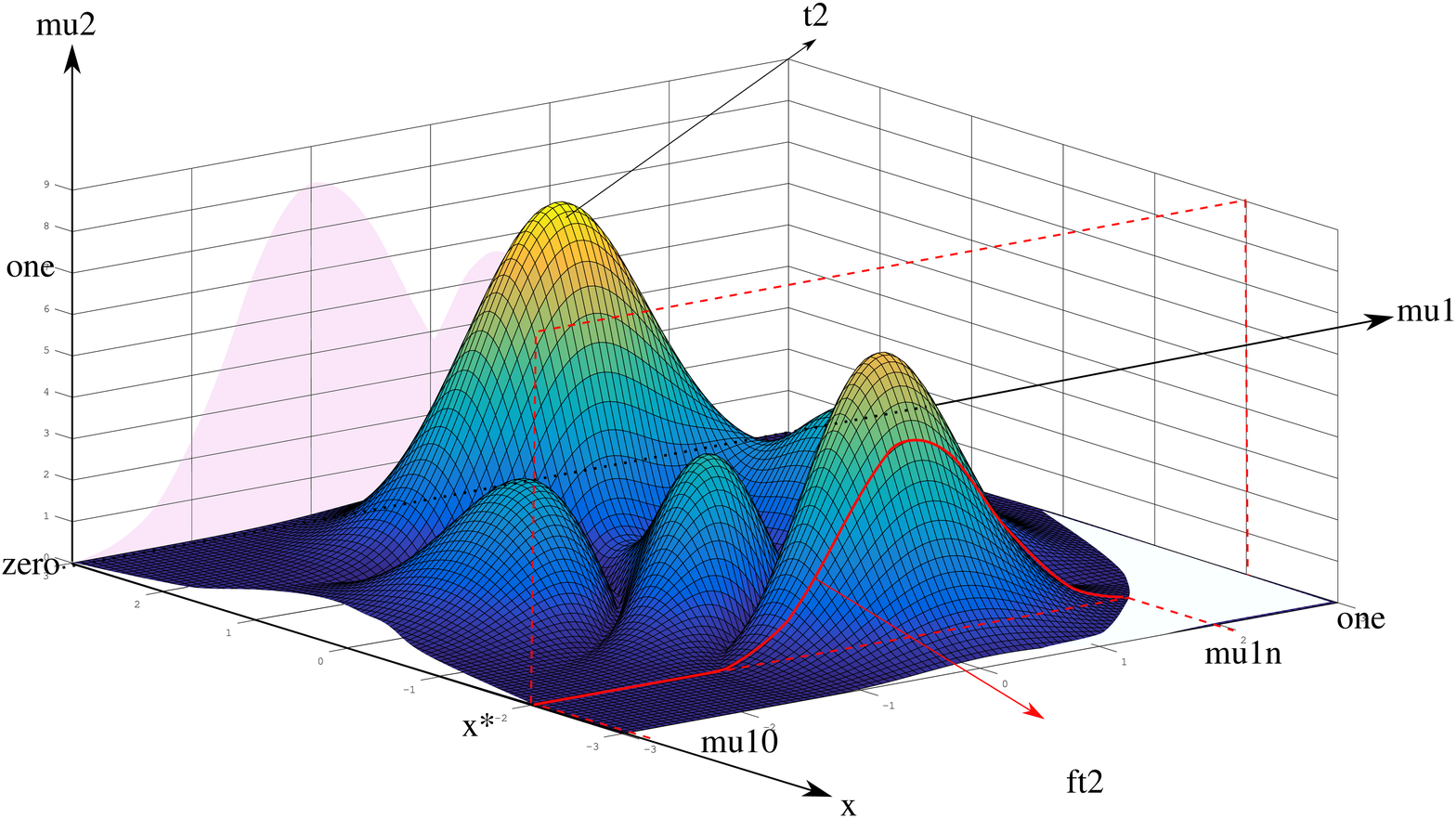}
\caption{Type-2 membership function (continuous domain).}
\label{fig:IEEE_Type2}
\end{center}
\end{figure} 

Type-2 (and higher) fuzzy sets \cite{Zadeh:1975} have been defined as an extension to 
type-1 fuzzy sets, to handle the uncertainties that may exist in the membership degrees themselves. 
Correspondingly, type-2 membership functions assign a set of $n$ values (in the limit $n \rightarrow \infty$) 
instead of a single one to the \emph{primary} membership degree of a point. 
Each primary membership degree $\fpone_i(x)$ of point $x$ adopts a \emph{secondary} membership degree  
$\fsone\left(x,\fpone_i(x)\right)$ between 0 and 1. 
Type-2 membership functions may be illustrated in a 3-dimensional space. 
Figure~\ref{fig:Primary_Secondary_MV} illustrates a discrete-time type-2 membership function, where 
point $x^*$ adopts three primary membership degrees $\mu_{1,1}$, $\mu_{1,2}$, and $\mu_{1,3}$, each  
corresponding to a secondary membership degree $\mu_{2,1}$, $\mu_{2,2}$, and $\mu_{2,3}$, respectively. 
Figure~\ref{fig:IEEE_Type2} shows a continuous-domain type-2 membership function with its secondary type-1 
membership functions, $\fsone (x,\cdot)$ defined for an arbitrary point $x$ (i.e., cross section of 
the 3-dimensional type-2 fuzzy membership function with a plane  parallel to the $\mu_1-\mu_2$ plane 
passing through point $x$).%

\subsection{Uncertainties: Probability versus fuzziness}
\label{sec:intro:uncertainties}

In this section, we shortly discuss the possible natures of uncertainties that may occur: 
\emph{probabilistic} and \emph{fuzzy} \cite{Eisen:1969}.%
  
A probabilistic uncertainty involves a set of random  events, represented by logical statements, 
each possessing a \emph{quantitative} expression and a certain probability ($\leq 1$) of occurrence. 
Based on probability theory \cite{Eisen:1969}, the summation of these probabilities for all the random events 
is one. 
For a set of random events, the uncertainty is in the possibility of occurrence of each random event. 
Probability functions may be used to describe a set of random events.%

A fuzzy uncertainty occurs due to the use of \emph{qualitative} expressions in a logical statement that 
can have various quantitative interpretations. Within a set of fuzzy events, each fuzzy event corresponds 
to a membership degree less than or equal to one. 
Due to the different interpretations of fuzzy statements, there may be overlaps in the quantitative interpretations 
of the fuzzy events in a set (i.e., realization of a fuzzy event does not necessarily exclude realization of other 
fuzzy events in the set). 
Consequently, the summation of the membership degrees of all the fuzzy events may exceed or be lower than one.
For instance, the three logical statements ``The room climate is cold'', ``The room climate is moderate'', 
``The room climate is warm'', represent fuzzy events, because the qualitative terms \emph{cold,} \emph{moderate}, 
and \emph{warm} may be linked to different temperature and humidity ranges when interpreted quantitatively. 
Fuzzy events can be modeled using fuzzy sets.%

\section{Novel concepts in type-2 fuzzy rules}
\label{sec:Novel_formulation}

In this section, we introduce the novel concept of probabilistic-fuzzy membership functions. 
Moreover, we propose a general formulation for type-2 nonlinear fuzzy rules.%

\subsection{Probabilistic-fuzzy and fuzzy-fuzzy membership functions}
\label{sec:ProbFuzzy_FuzzyFuzzy_MF}

Now, we look further at type-2 membership functions, and extend this concept based on the 
two types of uncertainties discussed in Section~\ref{sec:intro:uncertainties}. 
In order to cover both types of uncertainties, we expand the current concept of type-2 membership functions to two variants, 
which we call probabilistic-fuzzy and fuzzy-fuzzy membership functions.%

In case multiple uncertainties are integrated in a logical statement, fuzzy sets and membership functions of types 
higher than one may provide a higher accuracy in the fuzzification.  
We focus on type-2 fuzzy sets and membership functions (i.e., cases where two sources of uncertainty appear 
in a logical statement), while generalization to higher types can be done following a similar approach. 
We introduce two forms of integrated events: random-fuzzy and fuzzy-fuzzy.%
 
A random-fuzzy event is described by a logical statement that includes one quantitative and 
one qualitative term. 
The following three logical statements build up a set of random-fuzzy events: 
``The room climate is $20\%$ cold'', ``The room climate is $54\%$ moderate'', 
``The room climate is $26\%$ warm''. 
The first descriptive term for the room climate in each statement ($20\%$, $54\%$, $26\%$) 
represents a random event with a certain probability, 
while the second descriptive term (cold, moderate, warm) includes a qualitative term 
that can be interpreted and quantified in more than one way. 
Such logical statements that represent random-fuzzy events, can efficiently be modeled 
by a probabilistic-fuzzy membership function, which is a type-2 membership function with the 
primary membership function a type-1 membership function and the secondary membership function 
a probability function.%

A fuzzy-fuzzy event has a descriptive statement that involves two
qualitative terms. In the following
three logical statements: ``The room climate is slightly warm'', ``The
room climate is moderately warm'', ``The room climate is very warm'',
in addition to the qualitative term \emph{warm}, the terms
\emph{slightly}, \emph{moderately}, and \emph{very} can also have
several quantitative interpretations.  Such logical statements make a
set of fuzzy-fuzzy events that can be modeled using a fuzzy-fuzzy
membership function, where both the primary and the secondary
membership functions are type-1 membership functions.%

\subsection{Type-2 nonlinear fuzzy rules for input-delayed systems}
\label{sec:type2_nonlinear_fuzzyModel}

\begin{figure}
\begin{center}
\psfrag{k-1}[][][.7]{$\kz-1$}
\psfrag{k}[][][.7]{$\kz$}
\psfrag{k+1}[][][.7]{$\kz+1$}
\psfrag{k+2}[][][.7]{$\kz+2$}
\psfrag{k+3}[][][.7]{$\kz+3$}
\psfrag{k+4}[][][.7]{$\kz+4$}
\psfrag{k+5}[][][.7]{$\kz+5$}
\psfrag{k+6}[][][.7]{$\kz+6$}
\psfrag{k+7}[][][.7]{$\kz+7$}
\psfrag{discrete time k}[][][.7]{\hspace{3.5ex} discrete time $k$}
\includegraphics[width=0.8\linewidth]{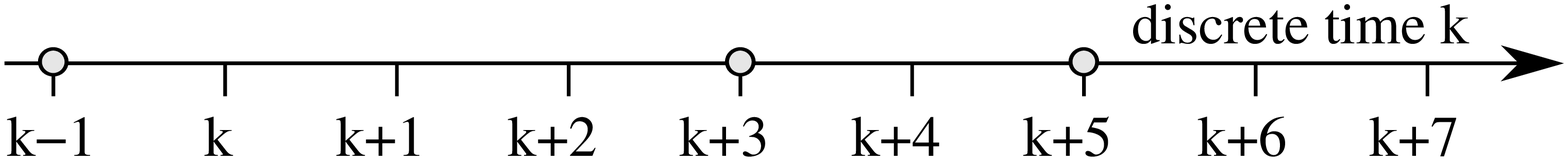}
\caption{The most reliable measurement available at various control time steps: 
$\pz{\kz}=\pz{\kz+1}=\pz{\kz+2}=\kz-1$, 
$\pz {\kz+4 }= \kz+3$, and 
$\pz  {\kz+ 6} = \pz {\kz+ 7} = \kz+5$ 
(The circular symbols illustrate those control time steps at which a reliable measurement 
of the state variables exist, where the measurements are not necessarily periodically available).}
\label{fig:previous_measurement}
\end{center}
\vspace*{1ex}
\begin{center}
\psfrag{k}[][][.7]{$\kz$}
\psfrag{k-1}[][][.7]{$\kz-1$}
\psfrag{k-2}[][][.7]{$\kz-2$}
\psfrag{k-3}[][][.7]{$\kz-3$}
\psfrag{k-4}[][][.7]{$\kz-4$}
\psfrag{k-5}[][][.7]{$\kz-5$}
\psfrag{k-6}[][][.7]{$\kz-6$}
\psfrag{k-7}[][][.7]{$\kz-7$}
\psfrag{k-8}[][][.7]{$\kz-8$}
\psfrag{discrete time k}[][][.7]{\hspace{3.5ex} discrete time $k$}
\includegraphics[width=0.8\linewidth]{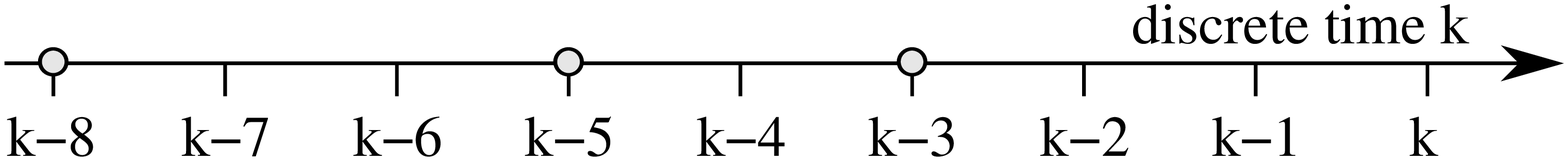}
\caption{The most reliable measurements (for $\delta=3$) at various control time steps: 
$\pz{\kz}=\pz{\kz-1}=\pz{\kz-2}=\kz-3$, 
$\pone{\kz} = \pone{\kz-1} = \pone{\kz-2}= \pz{\kz-4} = \kz-5$, and 
$\ptwo{\kz}=\ptwo{\kz-1}=\ptwo{\kz-2}=\pone{\kz-4}=\pz{\kz-6}=\pz{\kz-7}=\kz-8$.}
\label{fig:ctrl_prev_state_measurements}
\end{center}
\end{figure}

For some dynamical systems (including electric networks, pneumatic and hydraulic networks, chemical processes, 
long transmission lines \cite{Debeljkovic:2011}), at specific time steps only the time-delayed states of the system are available. 
This can be due to the long time span required for transferring the sensed data to the controller, 
slow sensors and measurement tools, costly measurement procedures and tools, missing or faulty measurements, 
or limited memory. 

In this section, we consider systems for which at time step $k$, the
delayed state variable, $\xm\left(\pz{k}\right)$, measured at time
step $\pz{k}$ (where $\pz{k}<k$), is just received or is the most
recent reliable measurement available (see
Figure~\ref{fig:previous_measurement}).  The dynamics of such systems
at time step $k$ should be formulated as a function of
$\xm\left(\pz{k}\right)$ and the control inputs that have affected the
system's dynamics from time step $\pz{k}$ until $k-1$.  We use $\xm$
to denote the \emph{measured state} and $\xe$ for the
\emph{estimated state} (by a model of the system).  A logical
``if-then'' rule for modeling the system's dynamics at time step $k$,
assuming there are no external disturbances, can in
general be stated as
\begin{align}
\label{eq:time_delayed_state_fuzzy_rule}
&\text{{\small if}}\hspace*{.6ex} \xm\bigl(\pz{k}\bigr)\in X_{\pz{k}}
\hspace{.15ex} \land \hspace*{.15ex} 
\bm{u}\left(\pz{k}\right) \in U_{\pz{k}}
\hspace*{.15ex} \land  \hspace*{.15ex}\ldots  
\bm{u}\left(k-1\right) \in U_{k-1},\nonumber\\
&\text{\small then} \hspace{.6ex} \xe\left(k\right)
=\fx\bigl(\tx(k),\xm\left( \pz{k} \right), \bm{u}\left( \pz{k}\right),\ldots,\bm{u}\left(k-1\right)\bigr),
\end{align}
where $k\in\Kc$ is the control step counter with $\Kc$ 
the set of all control time steps, 
$\pz{k}\in K^{\rm m}$ with $\Km$ the set of all discrete time steps at which a reliable measurement 
of the state variable exists, $\bm{u}$ is the control input vector, 
$X_{\pz{k}}$, $U_{\pz{k}}$, \ldots, $U_{k-1}$ are (generally fuzzy) sets,    
$\fx\left(\cdot\right)$ is a (generally nonlinear) function, and   
$\tx$ is a vector of design parameters.%

In such cases, where the current state variable is either unavailable or the reliability and accuracy of the 
realized measured value is questionable, a control system might be more robust when it 
considers multiple prior state variables, as well as the most recent one in making the current control decision. 
Correspondingly, we propose the following logical ``if-then'' rule for generating such control inputs:
\begin{align}
\label{eq:state_delayed_control_input_fuzzy_rule}
&\text{\small if}\hspace{.6ex} \xe  \bigl(k\bigr)\in X_{k}
\hspace*{.15ex} \land \hspace*{.15ex} 
\xm\bigl(\pz{k}\bigr) \in X_{\pz{k}}
\hspace*{.05ex} \land  \hspace*{.1ex} \ldots  
\xm  \bigl(\pdelta{k}\bigr) \in X_{\pdelta{k}},\nonumber\\
&\text{\small then}\\ 
&\hspace{.6ex} \bm{u}\left(k\right)
=\fu\bigl(\bm{\tu}(k),\xe \bigl(k\bigr),\xm \bigl(\pz{k}\bigr),\ldots,
\xm \bigl( {\pdelta{k}}\bigr), \kappa\left(k-1\right)\bigr), \nonumber
\end{align}  
where $X_{i}$ for $i\in\left\{\pdelta{k},\ldots,\pz{k},k\right\}$ are (generally fuzzy) sets, 
$\delta + 1$ is the number of previous state measurements involved, 
$\pi_{\mathsmaller{j}}\left(k\right)\in \Km$ for $j\in\left\{ 1,\ldots , \delta \right\}$ is the 
$j^{\rm th}$ most recent control time step prior to $\pz{k}$ (see Figure~\ref{fig:ctrl_prev_state_measurements} 
for an example) at which a reliable measurement of the state variable exists,  
$\fu\left(\cdot\right)$ is a (generally nonlinear) function, 
$\tu$ is a vector of design parameters,  
and $\kappa\left(k-1\right)$ is the average step cost value. 
\begin{remark}
The  last argument $\kappa\left(k-1\right)$ of function $\fu\left(\cdot\right)$ in  
\eqref{eq:state_delayed_control_input_fuzzy_rule} has been added  
to keep track of the cost value, and to prevent the resulting average cost value at the upcoming 
time step to grow significantly w.r.t.\ the previous time steps. 
\end{remark}
\smallskip

The value of $\kappa$ at time step $k$ is determined by
\begin{equation}
\label{eq:average_step_cost}
\kappa\left(k\right)= \frac{\dps 1}{\dps k+1} \sum_{l= 0}^{k} \lambda^{k-l} \hat{J}\left(\xe\left(l\right)
,\bm{u}\left(l\right), \bm{u}\left(l -1 \right)\right),
\end{equation}   
where $\hat{J}\left(\cdot\right)$ is the step cost function, i.e., a function that determines 
the realized value of the cost within one control sampling time, 
and $0<\lambda\leq 1$ is the \emph{forgetting factor}.
In order to reduce the required storage space for computation of the average step cost value, we 
propose the following updating equation for $\kappa$, which is derived from 
\eqref{eq:average_step_cost}: 
\begin{equation}
\label{eq:update_average_cost}
\kappa\left(k\right) = \frac{\dps k\lambda}{\dps k+1}\kappa\left(k-1\right) + 
                       \frac{\dps 1}{\dps k+1}\hat{J}\left(\xe\left(k\right)
                       ,\bm{u}\left(k\right), \bm{u}\left(k-1\right)\right).
\end{equation}
\begin{remark}
In \eqref{eq:average_step_cost}, we assume that $\bm{u}\left(-1\right) = 0$. 
\end{remark}
\begin{remark}
In case at some control time step, a reliable measured value of the state variable is received, 
$\xe$ in \eqref{eq:state_delayed_control_input_fuzzy_rule}
--\eqref{eq:update_average_cost} can be substituted by $\xm$ at that time step. 
\end{remark}
\begin{remark}
\label{remrak:combined_fuzzy_model}
A fuzzy model of the system has a fuzzy rule base consisting of several fuzzy rules 
of the form \eqref{eq:time_delayed_state_fuzzy_rule}. 
Each rule may produce a different value for a state variable. 
The final value can be obtained via a smooth   
linear combination of the values produced by all the rules (see \cite{Takagi:1985} for more details). 
For the sake of conciseness, we avoided adding an extra subscript $r$ 
to $\xe\left(k\right)$, $X_{\pz{k}}$,  $U_{\pz{k}}, \ldots, U_{k-1}$ , $\fx(\cdot)$, and $\tx(k)$ 
in \eqref{eq:time_delayed_state_fuzzy_rule}, as the counter of fuzzy rules.              
\end{remark}

\section{Two-layer predictive and multi-agent model-based intelligent control architecture}
\label{sec:two_layer_controller}

\begin{figure}
\begin{center}
\psfrag{mpc}[][][.85]{MPC module}
\psfrag{fuzzy}[][][.85]{Fuzzy-control module}
\psfrag{subsystem}[][][.85]{System}
\psfrag{first}[][][.85]{Bottom layer: Decision making}
\psfrag{second}[][][.85]{Top layer: Tuning and coordination}
\includegraphics[width=.28\textwidth]{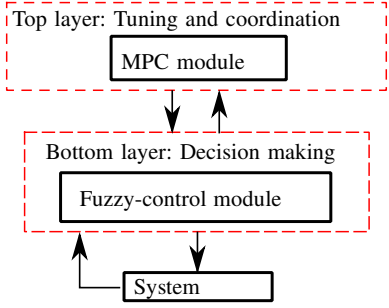}
\caption{The proposed two-layer control architecture.} 
\label{fig:Two_layer_architecture}
\end{center}
\end{figure}

In this section, we propose an integrated two-layer multi-agent control architecture that aims to minimize 
an overall cost value for large-scale and/or complex-dynamics systems with time-delayed, missing, or faulty measurements 
of the state variables. 
Figure~\ref{fig:Two_layer_architecture} illustrates a simplified version of the proposed 
control architecture. 
The bottom layer is directly connected to the actuators of the controlled 
system, and includes the intelligent-control module.%

The intelligent-control module may embed several distributed intelligent-control agents 
(we particularly use model-based type-2 fuzzy-control agents that are built upon the rules introduced in Section~\ref{sec:type2_nonlinear_fuzzyModel}). 
These fuzzy-control agents have predefined local cost functions and correspond to different subsystems, 
the dynamics of which may not be completely isolated from one another.  
Consequently, the control input computed by an agent for its subsystem may affect the dynamics and 
the cost value of other dynamically connected subsystems. 
This requires those fuzzy-control agents that are assigned to subsystems with connected dynamics to 
coordinate their decisions, such that the mutual effects of the control decisions on the state and cost 
values, do not result in any negative effects on the overall performance of the entire system. 
To that aim, the top control layer tunes the adaptive parameters of the fuzzy-control 
agents using an MPC module, such that the effects of the interactions of the dynamically 
connected subsystems are involved.%

The MPC optimization problem can be solved via a decomposition method, 
where the optimal solution is used to tune the parameters of the type-2 fuzzy-control agents (see 
Section~\ref{sec:fuzzy_controller}). The control inputs of the subsystems determined by the MPC 
module are based on the global cost function. 
Hence, using these optimal inputs for (re-)tuning the design parameters of the fuzzy-control agents can 
add the mutual influences and interactions of the subsystems within the updated fuzzy rules.%

A significant advantage of using the proposed two-layer predictive and multi-agent fuzzy 
control architecture compared with a distributed MPC-based architecture is in the very low computation 
time and hence, high speed of the fuzzy controllers w.r.t.\ an MPC-based one (which requires 
solving an optimization problem online). On the other hand, a fuzzy controller considers the previous and 
current state variables, without looking into the future. 
The presence of an MPC module in the top layer of the proposed control architecture will 
guarantee that the future impacts of decided control inputs will be considered. 
Moreover, to reduce the computational burden, the MPC module will solve the corresponding 
optimization problem only at the control time steps when it is called by the bottom control layer.%

\section{Bottom layer: Fuzzy-control module}
\label{sec:first_control_layer}

In this section, we explain the different elements within the bottom layer of the proposed integrated architecture.%

\subsection{Fuzzy model of a subsystem}
\label{sec:fuzzy_models}

For modeling the subsystems, we use the most recent reliable measured state and the corresponding control 
inputs from the time step of this measurement on, within the proposed formulation 
\eqref{eq:time_delayed_state_fuzzy_rule}. We consider general MIMO subsystems:  
\begin{align}
\label{eq:subsystem_fuzzy_model}
&\text{\small if}\hspace{.6ex} \xm_{i,s}\bigl(\pz{k}\bigr)\in X_{i,s,\pz{k}}
\nonumber\\ 
&\quad\bigwedge_{\ell = 1}^{\Nu_{i,s}}\hspace*{1ex}   
\bigwedge_{q=0}^{ k-1-\pz{k} } \bm{u}_{\ell,i,s}\left(\pz{k}+q\right) 
\in U_{\ell,i,s,\pz{k}+q}\nonumber\\
&\quad \bigwedge_{q=0}^{ k-1-\pz{k} } \bm{\nu}_{i,s}\left(\pz{k}+q\right) 
\in N_{i,s,\pz{k}+q}, \qquad\text{\small then}\\ 
&\hspace{1ex} \xe_{i,s}\left(k\right)=
\fx_{i,s}\Bigl(\txcon_{i,s}(k),\xm_{i,s}
\left(\pz{k}\right), \bar{\bm{u}}_{i,s}\left(k-1\right), 
\bar{\bm{\nu}}_{i,s}\left(k-1\right)\Bigr),\nonumber
\end{align}
with $\xm_{i,s}$ the measured value of the $i^{\rm th}$ state variable of subsystem $s$ 
(for $s\in\left\{1,\ldots,\ns\right\}$ with $\ns$ the total number of subsystems), 
$\xe_{i,s}$ the estimated value by the fuzzy model for the $i^{\rm th}$ state variable of 
subsystem $s$, $\pz{k}\in\Km$ the most recent control time step from $k$ at which a reliable 
measurement of the $i^{\rm th}$ state variable is available,  
$\Nu_{i,s}$ the number of control input vectors that affect the $i^{\rm th}$ state variable,  
$\bm{u}_{\ell,i,s}$ for $\ell\in\left\{1,\ldots,\Nu_{i,s}\right\}$ the $\ell^{\rm th}$ 
control input that influences the $i^{\rm th}$ state variable, 
$\bm{\nu}_{i,s}$ the external disturbance that corresponds to the $i^{\rm th}$ state variable, 
$X_{i,s,\pz{k}}$, $U_{\ell,i,s,\pz{k}+q}$, and $N_{i,s,\pz{k} +q}$ for 
$q\in\left\{0,\ldots,k-1-\pz{k}\right\}$ (generally type-2 fuzzy) sets,  
$\bar{\bm{u}}_{i,s}\left(k-1\right)$ and $\bar{\bm{\nu}}_{i,s}\left(k-1\right)$  
vectors with, respectively, $\left(k-\pz{k}\right)\cdot  \Nu_{i,s}$ 
elements of $\bm{u}_{\ell,i,s}$ and $k-\pz{k}$ elements of $\bm{\nu}_{i,s}$ from  
control time step $\pz{k}$ until control time step $k-1$, 
$\fx_{i,s}\left(\cdot\right)$ a generally nonlinear function, 
and $\txcon_{i,s}(k)$ a vector including all the adaptive parameters of the consequent 
of the fuzzy rule  at control time step $k$.  
\smallskip
\begin{remark}
The vector $\txcon_{i,s}(k)$ will be (re-)identified at specific identification time steps 
$k\in\Kid_s$ (which do not necessarily coincide with every control time step) in order to 
update the fuzzy model and make it more accurate based on the most recent information on   
the dynamics of the system (see Section~\ref{sec:parameter_identification} for more details). 
\end{remark}
\begin{remark}
\label{rem:Kid}
The set $\Kid_s$ is constructed on a mixed, regular and
event-triggered basis.  The set includes some preset time steps at
which a measurement of the state variables of subsystem $s$ is
supposed to be available.  Additionally, a set of identification
thresholds is defined based on the model's estimation error w.r.t.\
the realized values of the measurements. In the real-time run of the
model, in case at least one of the identification thresholds is
exceeded at a specific time step, that time step will be added to
$\Kid_s$ and the parameters of the fuzzy model are re-identified.
\end{remark}
\begin{remark}
Some adaptive parameters can be considered in the mathematical formulation of the 
sets $X_{i,s,\pz{k}}$, $U_{\ell,i,s,\pz{k}+q}$, and $N_{i,s,\pz{k}+q}$, which are stored 
in a vector denoted by $\txant_{i,s}(k)$. This vector may also be 
updated together with $\txcon_{i,s}(k)$ at the identification time steps $k\in\Kid_s$.   
\end{remark}

\subsection{Fuzzy-control agent of a subsystem}
\label{sec:fuzzy_controller}

Each fuzzy-control agent in the bottom layer of control uses the following 
adaptive state and cost-feedback rule,  
inspired by \eqref{eq:state_delayed_control_input_fuzzy_rule}, 
to steer the actuators of its subsystem:  
\begin{align}
\label{eq:agent_fuzzy_controller}
\text{\small if}\hspace*{.6ex} 
&\bigwedge_{\ell=1}^{\nx_{i,s}}\left[
\xe_{\ell,i,s}\left(k\right)\in X_{\ell,i,s,k}
\bigwedge_{q\in\left\{\pdelta{k},\ldots,\pz{k}\right\}}
\xm_{\ell,i,s}\left(q\right)\in X_{\ell,i,s,q}
\right]
\nonumber\\ 
& \bigwedge_{\ell=1}^{\nx_{i,s}}\hspace*{1ex} \bigwedge_{q\in\left\{\pdelta{k},\ldots,\pz{k},k\right\}}
\bm{\nu}_{\ell,i,s}\left(q\right) \in N_{\ell,i,s,q},\\
&\text{\small then}
\hspace*{1ex}\bm{u}_{i,s}\left(k\right)
=\fu_{i,s}\Bigl(\tucon_{i,s}\left(k\right),\bar{\bm{x}}_{i,s,\delta}\left(k\right), 
\bar{\bm{\nu}}_{i,s,\delta}\left(k\right), \kappa_s\left(k-1\right) \Bigr),\nonumber
\end{align}
where $\xe_{\ell,i,s}$ and $\xm_{\ell,i,s}$ for $\ell\in\left\{1,\ldots,\nx_{i,s}\right\}$ are the 
estimated and the measured values of the $\ell^{\rm th}$ state variable 
of the MIMO  subsystem $s$ that is influenced directly by the control input $\bm{u}_{i,s}$, 
$\nx_{i,s}$ is the total number of such state variables,  
$\bm{u}_{i,s}$ is the $i^{\rm th}$ control input of subsystem $s$, 
$\pi_i\left(k\right)\in\Km$ for $i\in\left\{0,\ldots,\delta\right\}$ is the $i^{\rm th}$ most 
recent control time step from the current time step $k$ at which a reliable measurement 
of the $\ell^{\rm th}$ state variable that is directly affected by $\bm{u}_{i,s}$ exists, 
$\delta+1$ is  the total number of the previous time  
steps the measurements $\xm_{\ell,i,s}$ of which are used by the fuzzy-control agent,   
$\bm{\nu}_{\ell,i,s}$ includes all the external disturbances that correspond to  the $\ell^{\rm th}$ state variable 
that is directly influenced by $\bm{u}_{i,s}$, $X_{\ell,i,s,q}$ and $N_{\ell,i,s,q}$ for $q\in\left\{\pdelta{k},\ldots,\pz{k},  k\right\}$ 
are (generally fuzzy) sets, $\fu_{i,s}\left(\cdot\right)$ is a generally nonlinear function, 
$\tucon_{i,s}$ is a vector that consists of all parameters of the consequent of the fuzzy rule, 
and $\bar{\bm{x}}_{i,s,\delta} (k) $ and $\bar{\bm{\nu}}_{i,s,\delta}(k)$ are vectors that include, respectively, 
all the state variables $\xm_{\ell,i,s}$ form control time step $\pdelta{k}$ 
until control time step $k$ (for the current time step, $\xe_{\ell, i, s}\left(k\right)$ may be used 
instead, if the measurement is not yet available) and the corresponding external disturbances. 
The average step cost value $\kappa_s$ for subsystem $s$ at control time step $k-1$ is computed by 
the approach explained in Section~\ref{sec:type2_nonlinear_fuzzyModel} 
(see \eqref{eq:average_step_cost} and \eqref{eq:update_average_cost}).%
\smallskip
\begin{remark}
The vector $\tucon_{i,s}(k)$ will be (re-)tuned at specific tuning time steps $k\in\Ktune_s$ 
in order to improve the performance of the resulting fuzzy control rules adaptively and based 
on the most recent information on the dynamics of the controlled system 
(see Section~\ref{sec:ctrl_parameter_tuning} for more details). 
\end{remark}
\begin{remark}
\label{remark:Ktune}
Similarly to $\Kid_s$, the  set $\Ktune_s$ is constructed based on a mixed, regular and event-triggered 
approach. 
The set includes some preset regular control  time steps, at which the tuning module in the top control  
layer will be activated.  
Additionally, in case some predefined tuning criteria are triggered (e.g., if the most recent realized value 
of the step cost function exceeds a threshold), re-tuning of the controller parameters occurs.%
\end{remark}%
\begin{remark}
Some adaptive parameters may be considered in the mathematical formulation of the sets 
$X_{\ell,i,s,q}$ and $N_{\ell,i,s,q}$. These parameters will be stored in the vector $\tuant_{i,s}$, 
which may be updated together with $\tucon_{i,s}(k)$ at the tuning time steps $k\in\Ktune_s$.   
\end{remark}

\subsection{Parameter identification for fuzzy models}
\label{sec:parameter_identification}

\begin{figure}
\begin{center}
\psfrag{system}[][][.85]{Subsystem $s$}
\psfrag{model}[][][.85]{Fuzzy model $s$}
\psfrag{p}[][][.7]{$\bm{+}$}
\psfrag{m}[][][.7]{$\bm{-}$}
\psfrag{x}[][][.85]{${\color{Cyan}\xm_s}$}
\psfrag{xe}[][][.85]{${\color{Cyan}\xe_s}$}
\psfrag{e}[][][.85]{${\color{Cyan}{{\bm \epsilon}}^{\rm e}}$}
\psfrag{evaluation}[][][.85]{Evaluation and minimization of}
\psfrag{minimization}[][][.85]{cumulative estimation error}
\psfrag{tconp}[][][.7]{${\color {ForestGreen}\txcon_{i,s}}$}
\psfrag{tcon}[][][.7]{${\color {ForestGreen}\bm{\theta}^{{\bm x}, {\rm con, up}}_{i,s}}$}
\includegraphics[width=.65\linewidth]{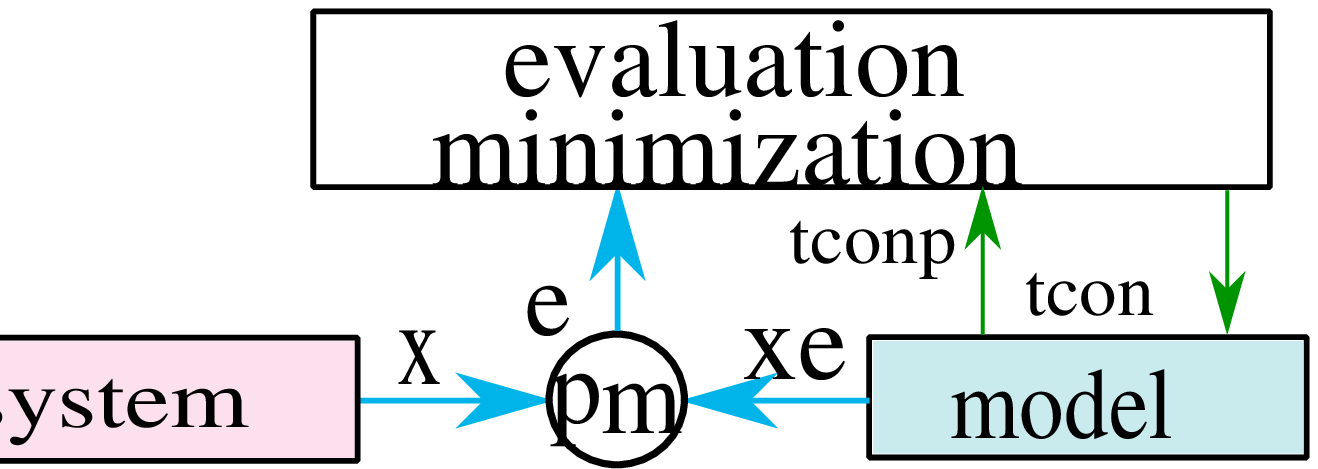}
\caption{Identifying the parameters of the fuzzy model of subsystem $s$.}
\label{fig:simplified_identification}
\end{center}

\vspace*{1ex}

\begin{center}
\psfrag{dis}[][][.85]{Disturbance}
\psfrag{storage}[][][.85]{storage}
\psfrag{2L}[][][.85]{Two-layer}
\psfrag{ctrl}[][][.85]{control system}
\psfrag{model}[][][.85]{Fuzzy model $s$}
\psfrag{ctrl storage}[][][.85]{Control input storage}
\psfrag{m storage}[][][.85]{Measurement storage}
\psfrag{system}[][][.85]{Subsystem~$s$}
\psfrag{model evaluator}[][][.85]{Model evaluator}
\psfrag{identification module}[][][.85]{Identification module:}
\psfrag{re-identification}[][][.85]{(Re-)identifying antecedent 
and consequent parameters via \eqref{eq:Model_antecedent_identification} 
and \eqref{eq:Model_consequent_identification}}
\psfrag{x}[][][.85]{${\color{Cyan}\xm_s}$}
\psfrag{u}[][][.85]{$\bm{u}_s$}
\psfrag{up}[][][.85]{$\bm{u}^{\rm p}_s$}
\psfrag{xp}[][][.85]{$\bm{x}^{\rm m,p}_s$}
\psfrag{v}[][][.85]{$\bm{\nu}_s$}
\psfrag{vp}[][][.85]{$\bm{\nu}^{\rm p}_s$}
\psfrag{xe}[][][.85]{${\color{Cyan}\xe_s}$}
\psfrag{p}[][][.7]{$\bm{+}$}
\psfrag{m}[][][.7]{$\bm{-}$}
\psfrag{X}[][][.7]{${\color {ForestGreen}X^{\rm{m}}_s}$}
\psfrag{e}[][][.85]{${\color{Cyan}{{\bm \epsilon}}^{\rm e}}$}
\psfrag{e>e}[][][.7]{$\bm{\epsilon}^{\rm e}\geq \bar{\bm \epsilon}\ \Rightarrow$ closed}
\psfrag{e<e}[][][.7]{$\bm{\epsilon}^{\rm e} < \bar{\bm \epsilon} \ \Rightarrow$ opened}
\psfrag{in Kid}[][][.7]{$k\in\Kid_s\ \Rightarrow$ closed}
\psfrag{not in Kid}[][][.7]{$k\not\in\Kid_s\ \Rightarrow$ opened}
\psfrag{tantp}[][][.7]{${\color {ForestGreen}\txant_{i,s}}$}
\psfrag{tconp}[][][.7]{${\color {ForestGreen}\txcon_{i,s}}$}
\psfrag{tant}[][][.7]{${\color {ForestGreen}\bm{\theta}^{{\bm x}, {\rm ant, up}}_{i,s}}$}
\psfrag{tcon}[][][.7]{${\color {ForestGreen}\bm{\theta}^{{\bm x}, {\rm con, up}}_{i,s}}$}
\includegraphics[width=\linewidth]{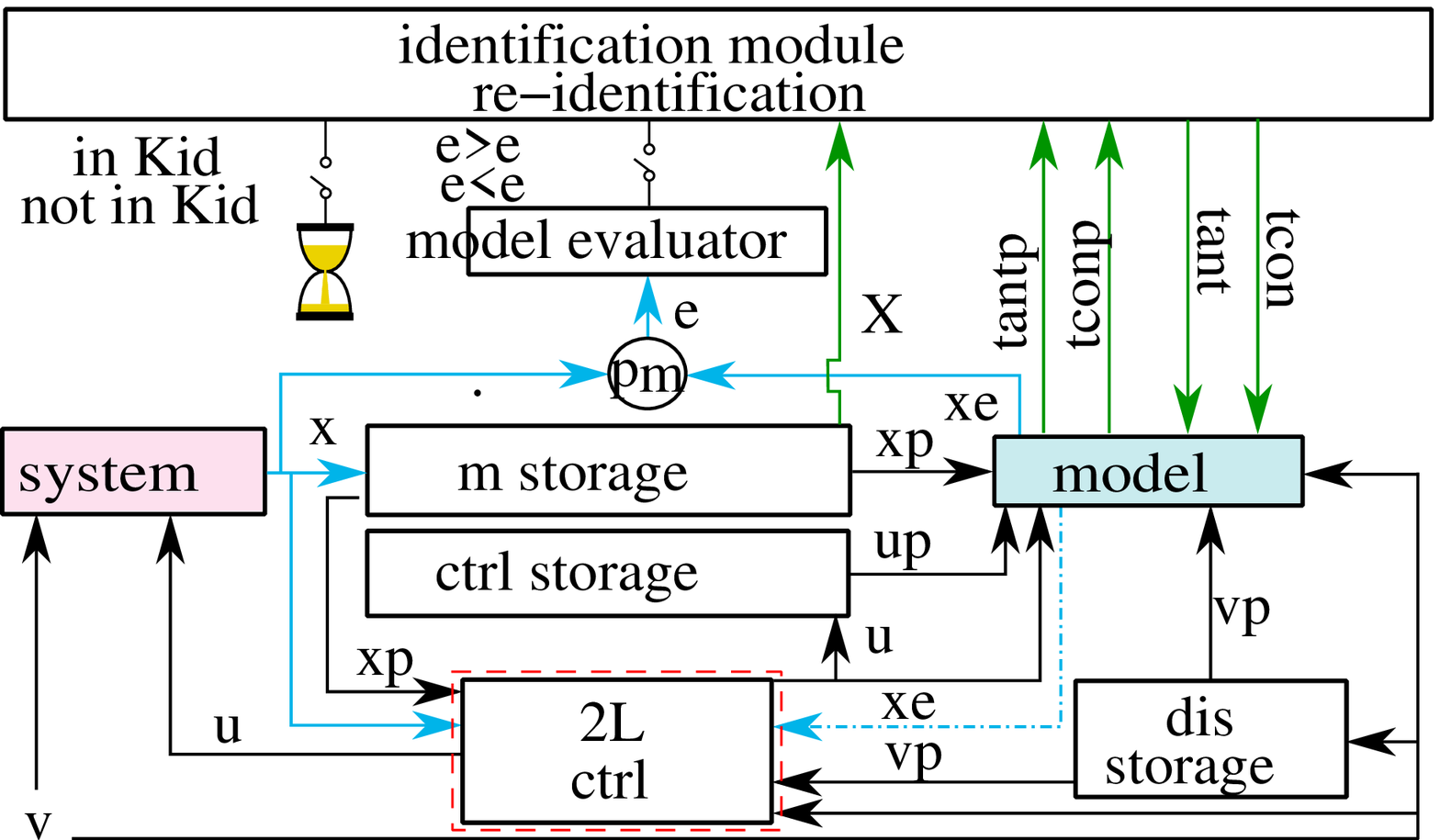}
\caption{The proposed two-layer control system: 
The blue solid signals are active at time steps when a measurement is received from 
the subsystem. 
The blue dash-dotted signal is active when the blue solid signals are not. 
The green signals demonstrating the model (re-)identification procedure are active 
at specific preset time steps $k\in\Kid_s$ and whenever the model evaluator 
recognizes based on the estimation error $\bm{\epsilon}^{\rm e}$ that an identification 
threshold $\bar{\bm \epsilon}$ has been exceeded. The superscripts ``p'' and 
``f''  stand for ``future'' and ``past'', respectively.}
\label{fig:subsystem_model}
\end{center}
\end{figure}

For the fuzzy model of every subsystem $s$ (a collection of fuzzy rules with the formulation  
\eqref{eq:subsystem_fuzzy_model} defined for all the state variables of the subsystem), 
the parameter vectors $\txant_{i,s}(k)$ and $\txcon_{i,s}(k)$ including the antecedent 
and consequent parameters of the fuzzy rules, are (re-)identified at time steps $k\in\Kid_s$.    
The most recent element of $\Kid_s$ before the current time step $k$ is shown by $\pid\left(k\right)$. 
Re-identification of $\txant_{i,s}$ at control time step $k$ is influenced by  
the most recent identified vector, $\txant_{i,s}\left(\pid\left(k\right)\right)$, 
the datasets $\bar{X}_{i,s,\delid}(k)$ and $\bar{N}_{i,s,\delid}(k)$, which consist of the elements of vectors 
$\bar{\bm{x}}_{i,s,\delid}(k)$ and $\bar{\bm{\nu}}_{i,s,\delid}(k)$ 
(with $\delid$ the number of reliable measurements of the states and external disturbances 
received at the sampling times between $\pid\left(k\right)$ and $k$), 
and dataset $\bar{U}^{\rm opt}_{i,s,\delid}(k)$, which includes the optimal values of the control input 
that influences the $i^{\rm th}$ state of subsystem $s$ directly, between  time steps $\pid\left(k\right)$ and $k$. 
The elements of $\bar{U}^{\rm opt}_{i,s, \delid} (k)$ can be determined offline using fast multi-parametric 
optimization approaches \cite{Bemporad:2002}:
\begin{align}
\label{eq:Model_antecedent_identification}
\txant_{i,s}&\left(k\right)=\\
\bm{\hat{\theta}}^{\bm{x}}\Bigl( 
&\txant_{i,s}\left(\pid\left(k\right)\right),\bar{X}_{i,s,\delid}(k), 
\bar{U}^{\rm opt}_{i,s, \delid} (k), 
\bar{N}_{i,s,\delid}(k)
\hspace*{-.5ex}\Bigr),
\nonumber
\end{align}
with $\bm{\hat{\theta}}^{\bm{x}}\left(\cdot\right)$ a generally nonlinear operator. 
The updated type-2 fuzzy sets in the antecedents of the fuzzy rules of the subsystem's  
model are obtained by
\begin{align}
\Big\{
X_{i,s,\pz{k}};\  
&U_{\ell,i,s,\pz{k}} : U_{\ell,i,s,k-1}, \ \ell\in\left\{1,\ldots,\Nu_{i,s}\right\}; 
\nonumber\\
&N_{i,s,\pz{k}}:N_{i,s,k-1}\Big\}
=\hat{\pi}^{\bm{x}}\left(\txant_{i,s}\left(k\right)\right),\nonumber
\end{align}   
where $\hat{\pi}^{\bm{x}}(\cdot)$ is a generally nonlinear operator that receives the corresponding 
parameters and gives the type-2 fuzzy sets of the antecedent. Note that $a_1:a_n$ is used for the sake of 
brevity of the notations and is equivalent to $a_1,\ldots,a_n$.%

The parameter vectors $\txcon_{i,s}\left(k\right)$ of the fuzzy rules in the  
model of subsystem $s$ can be updated at time step $k\in\Kid_s$, by minimizing 
the cumulative error of the state variables estimated by the model within a predefined time window, 
w.r.t\ their measured values (see Figure~\ref{fig:simplified_identification}). 
This time window at control time step $k$ is denoted by $\Lid_s\left(k\right)$,  
and includes a predefined number of the most recent elements within $\Kid_s$. 
We can write   
\begin{align}
\label{eq:Model_consequent_identification}
&\min_{\txcon_{i,s}\left(k\right)} \left(\mathsmaller{\sum}\nolimits_{l \in \Lid_s\left(k\right) } 
\norm{\xm_{i,s}\left(l\right)-\xe_{i,s}\left(l \big| \txcon_{i,s}\left(l\right)=
\txcon_{i,s}\left(k\right) \right)}\right), \nonumber\\
&\text{\rm s.t. \eqref{eq:subsystem_fuzzy_model} and \eqref{eq:agent_fuzzy_controller} for 
$l\in \Lid_s\left(k\right)$}. 
\end{align} 
In \eqref{eq:Model_consequent_identification}, the fuzzy model \eqref{eq:subsystem_fuzzy_model}  
is re-run within the time window $\Lid_s\left(k\right)$, assuming that the updated $\txcon_{i,s}\left(k\right)$ 
at time step $k$ is used for all the previous time steps. 
The optimization problem \eqref{eq:Model_consequent_identification} is in general 
nonlinear, nonsmooth, and nonconvex, and can be solved by standard 
optimization algorithms, such as pattern search, genetic algorithm, or gradient-based optimization approaches, using multiple starting points.%

Figure~\ref{fig:simplified_identification} shows a simplified view of
the identification procedure for the consequent parameters of the
fuzzy model for subsystem $s$ (note that the updated version of the
vector of the consequent parameters 
is indicated by $\bm{\theta}^{{\bm x}, {\rm con, up}}_{i,s}$).  More
details, including
(re-)identification of 
the antecedent and consequent parameters and the interactions and
connections with the proposed two-layer control system, are
illustrated in Figure~\ref{fig:subsystem_model}.%
The two-layer control system, receives the time-delayed measurements 
(indicated by $\bm{x}^{\rm m,p}_s$)
from the measurement storage and the current measured state 
(indicated by $\bm{x}^{\rm m}_s$)
from the subsystem in case it is available, or otherwise the estimated
value of the current state variable (indicated by $\bm{x}^{\rm e}_s$)
provided by the fuzzy model, together with the current and the
previous corresponding external disturbances (indicated by
$\bm{\nu}_s$ and $\bm{\nu}^{\rm p}_s$).  
The control system then produces the control input $\bm{u}_s$ and
injects it to the subsystem and to the control input storage.  In
Figure~\ref{fig:subsystem_model}, the signals illustrated by solid
blue arrows correspond to those time steps at which a reliable
measurement of the subsystem's state variable exists.  The signal that
is shown by the dash-dotted blue arrow realizes otherwise.  From this
figure, we see that the identification procedure occurs at specific
time steps $k\in\Kid_s$ (cf.\
Remark~\ref{rem:Kid}).
The signals that are illustrated by solid green arrows will be
activated only at these identification time steps.


\subsection{Parameter tuning for fuzzy-control agents}
\label{sec:ctrl_parameter_tuning}

\begin{figure}
\begin{center}
\psfrag{system}[][][.85]{Subsystem~$s$}
\psfrag{fuzzy controller}[][][.85]{Fuzzy-control agent $s$}
\psfrag{retune}[][][.85]{Re-tuning fuzzy-control agent $s$}
\psfrag{sigma}[][][.85]{$\hat{J}_s\geq \sigma$}
\psfrag{xm}[][][.85]{${\color{Cyan}\xm_s}$}
\psfrag{u}[][][.85]{$\bm{u}_s$}
\psfrag{tant}[][][.7]{${\color {ForestGreen}\tuant_{i,s}}$}
\psfrag{tcon}[][][.7]{${\color {ForestGreen}\tucon_{i,s}}$}
\psfrag{tantup}[][][.7]{${\color {ForestGreen}\bm{\theta}^{{\bm u}, {\rm ant, up}}_{i,s}}$}
\psfrag{tconup}[][][.7]{${\color {ForestGreen}\bm{\theta}^{{\bm u}, {\rm con, up}}_{i,s}}$}
\psfrag{compute}[][][.85]{Compute and evaluate $\hat{J}_s$}
\includegraphics[width=.6\linewidth]{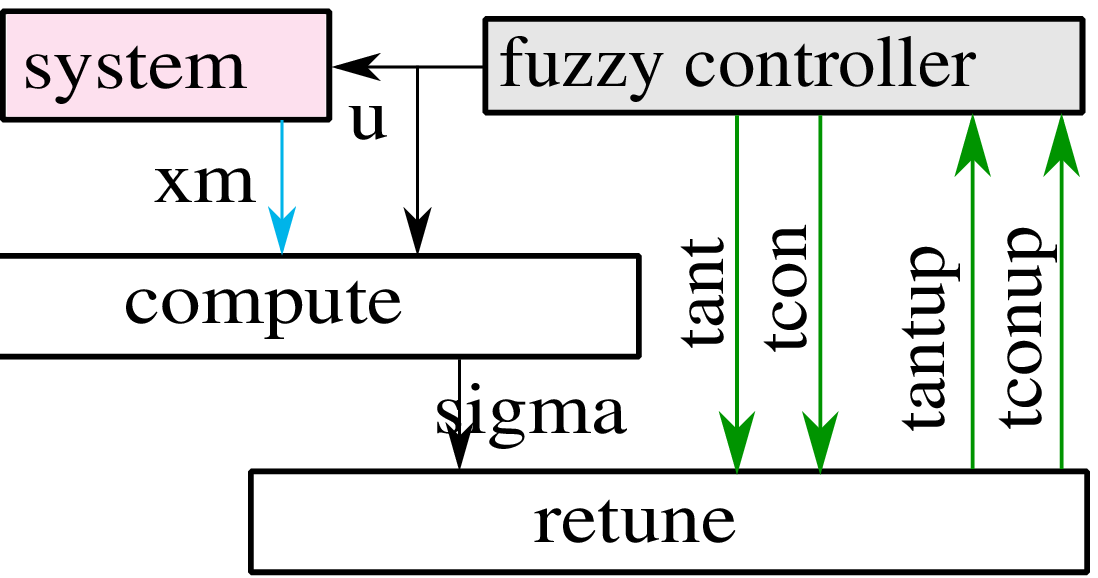}
\caption{Tuning the parameters of fuzzy-control agent $s$.}
\label{fig:SimplifiedTuning}
\end{center}

\vspace*{1ex}

\begin{center}
\psfrag{first layer}[][][.85]{{\color{red}Bottom layer}}
\psfrag{second layer}[][][.85]{{\color{red}Top layer}}
\psfrag{dis}[][][.85]{Disturbance}
\psfrag{storage}[][][.85]{storage}
\psfrag{2L}[][][.85]{Two-layer}
\psfrag{ctrl}[][][.85]{control system}
\psfrag{model 1}[][][.85]{Fuzzy model $1$}
\psfrag{model n}[][][.85]{Fuzzy model~$\ns$}
\psfrag{model}[][][.85]{Fuzzy model $s$}
\psfrag{ctrl storage}[][][.85]{Control input storage}
\psfrag{m storage}[][][.85]{Measurement storage}
\psfrag{system}[][][.85]{Subsystem~$s$}
\psfrag{fuzzy controller}[][][.85]{Fuzzy-control agent $s$}
\psfrag{jhat}[][][.85]{$\hat{J}_s$}
\psfrag{comp}[][][.85]{Computation module}
\psfrag{comp cost}[][][.85]{Computation module (cumulative cost)}
\psfrag{step cost}[][][.85]{(step cost)}
\psfrag{ctrl evaluator}[][][.85]{Control evaluator}
\psfrag{mpc}[][][.85]{MPC module}
\psfrag{ant tuning}[][][.85]{Antecedent tuning module}
\psfrag{via}[][][.85]{(tuning via \eqref{eq:Controller_antecedent_tuning})}
\psfrag{con tuning}[][][.85]{Consequent tuning module}
\psfrag{optimizer}[][][.85]{(Optimizer \eqref{eq:consequent_optimizing_module})}
\psfrag{phi1}[][][.85]{${\color {ForestGreen}\phi_1}$}
\psfrag{phin}[][][.85]{${\color {ForestGreen}\phi_{\ns}}$}
\psfrag{Theta}[][][.85]{${\color {ForestGreen}\phi_s}$}
\psfrag{x}[][][.85]{${\color{Cyan}\xm_s}$}
\psfrag{u}[][][.85]{$\bm{u}_s$}
\psfrag{up}[][][.85]{$\bm{u}^{\rm p}_s$}
\psfrag{uint}[][][.85]{${\color{ForestGreen}
\begin{bmatrix}
\bm{u}_1\\
\vdots\\
\bm{u}_s\\
\vdots\\
\bm{u}_{\ns}
\end{bmatrix}
}$}
\psfrag{ug}[][][.85]{${\color{ForestGreen}\bm{u}_s}$}
\psfrag{xp}[][][.85]{$\bm{x}^{{\rm m,p}}_s$}
\psfrag{v}[][][.85]{\hspace*{1ex}$\bm{\nu}_s$}
\psfrag{vp}[][][.85]{$\bm{\nu}^{{\rm p}}_s$}
\psfrag{xe}[][][.85]{${\color{Cyan}\xe_s}$}
\psfrag{xeg}[][][.85]{${\color{ForestGreen}\xe_s}$}
\psfrag{xint}[][][.85]{${\color{ForestGreen}\xe}$}
\psfrag{X}[][][.7]{${\color {ForestGreen}X^{\rm{m}}}$}
\psfrag{xc}[][][.7]{${\color {ForestGreen}\bm{x}^{\rm{c}}}$}
\psfrag{e}[][][.85]{${\color{Cyan}{{\bm \epsilon}}^{\rm e}}$}
\psfrag{j>sig}[][][.7]{$\hat{J}_s\geq \sigma\ \Rightarrow$ closed}
\psfrag{j<sig}[][][.7]{$\hat{J}_s < \sigma\ \Rightarrow$ opened}
\psfrag{in Kt}[][][.7]{$k\in\Ktune_s\ \Rightarrow$ closed}
\psfrag{not in Kt}[][][.7]{$k\not\in\Ktune_s\ \Rightarrow$ opened}
\psfrag{tant}[][][.7]{${\color {ForestGreen}\tuant_{i,s}}$}
\psfrag{tcon}[][][.7]{${\color {ForestGreen}\tucon_{i,s}}$}
\psfrag{tantup}[][][.7]{${\color {ForestGreen}\bm{\theta}^{{\bm u}, {\rm ant, up}}_{i,s}}$}
\psfrag{tconup}[][][.7]{${\color {ForestGreen}\bm{\theta}^{{\bm u}, {\rm con, up}}_{i,s}}$}
\psfrag{Jp}[][][.85]{${\color{ForestGreen}\Sigma_s\hat{J}_s^{\rm p}}$}
\psfrag{Jf}[][][.85]{${\color{ForestGreen}\Sigma_s\hat{J}_s^{\rm f}}$}
\psfrag{vp-g}[][][.85]{${\color {ForestGreen}\bm{\nu}^{{\rm p}}_s}$}
\psfrag{xp-g}[][][.85]{${\color {ForestGreen}\bm{x}^{{\rm m,p}}_s}$}
\includegraphics[width=\linewidth]{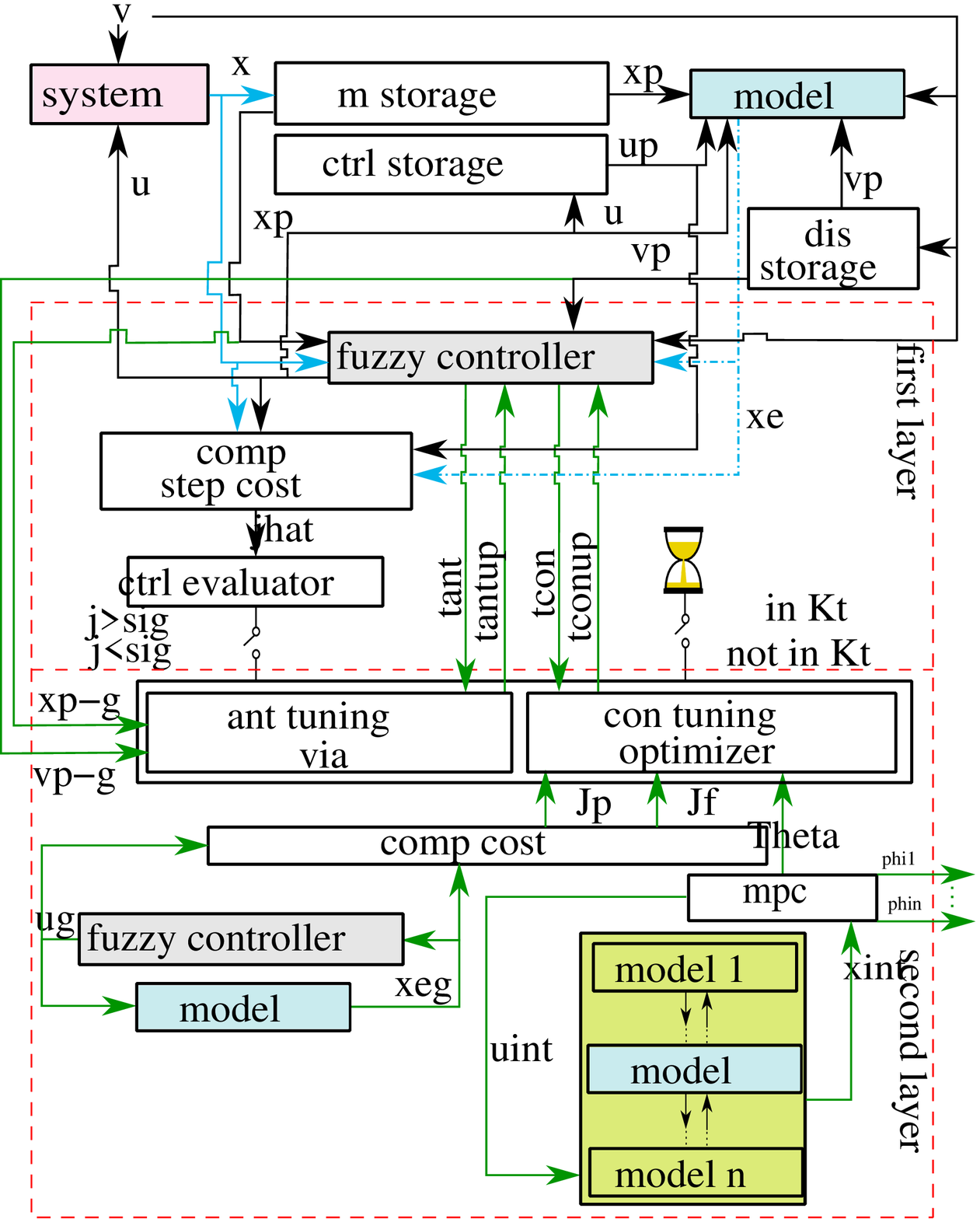}
\caption{The detailed architecture of the proposed two-layer control system: 
Parameter (re-)tuning occurs (i.e., the signals shown by the green arrows are activated) at time steps 
$k\in\Ktune_s$ (including specific preset time steps and time steps when a tuning criterion is triggered, 
i.e., the most recent value of the step cost exceeds a predefined threshold).}
\label{fig:two_layer_controller}
\end{center}
\end{figure}

A fuzzy-control agent specified by \eqref{eq:agent_fuzzy_controller}, includes two vectors 
$\tuant_{i,s}$ and $\tucon_{i,s}$ of design parameters that should be\linebreak (re-)tuned regularly 
(at time steps $k\in\Ktune_s$), to guarantee a satisfactory performance for the control system.  
The two-layer control architecture (see Figure~\ref{fig:Two_layer_architecture} for 
an overall view and Figure~\ref{fig:two_layer_controller} for extra details) has been designed 
specifically for efficient performance and tuning of the fuzzy controller \eqref{eq:agent_fuzzy_controller}. 
In order to clarify the links between Figures~\ref{fig:Two_layer_architecture}, \ref{fig:subsystem_model}, 
and \ref{fig:two_layer_controller}, the dashed red boxes have been used. 
The detailed structure shown within the red boxes  
in Figure~\ref{fig:two_layer_controller} includes the main elements of the proposed 
two-layer control system, which is shown and encountered by similar dashed red boxes in 
Figures~\ref{fig:Two_layer_architecture} and \ref{fig:subsystem_model}.       
Next, we elaborate the tuning procedure and Figure~\ref{fig:two_layer_controller}.%

\smallskip

Figure~\ref{fig:SimplifiedTuning} simplifies the tuning procedure, 
which occurs at the top control layer, while 
Figure~\ref{fig:two_layer_controller} shows the detailed structure of both layers of the 
proposed integrated control system.  
From Figure~\ref{fig:two_layer_controller}, at some regular and triggered control time steps 
(when the step cost value $\hat{J}_s$ exceeds a threshold $\sigma$), 
the top control layer comes into action. 
In that case, the signals shown in green in Figures~\ref{fig:SimplifiedTuning} and 
\ref{fig:two_layer_controller} will be activated.   
The parameters $\tuant_{i,s}$ of the antecedents of the fuzzy control rules 
for subsystem $s$ are updated based on the most recent values of the vector 
$\tuant_{i,s}$ updated at time step $\ptune\left(k\right)\in\Ktune$, and the datasets 
$\bar{X}_{i,s,\deltune}(k)$ and $\bar{N}_{i,s,\deltune}(k)$, 
which include the elements of $\bar{\bm{x}}_{i,s,\deltune}(k)$ and $\bar{\bm{\nu}}_{i,s,\deltune}(k)$  
(with $\deltune$ the number of the reliable measurements of the states and external disturbances 
available between control time steps $\ptune\left(k\right)$ and $k$). Therefore, we have    
\begin{align}
\label{eq:Controller_antecedent_tuning}
\tuant_{i,s}\left(k\right)=\bm{\hat{\theta}}^{\bm{u}}\Bigl(\tuant_{i,s}\left(\ptune\left(k\right)\right), 
\bar{X}_{i,s,\deltune}(k), \bar{N}_{i,s,\deltune}(k)
\Bigr),
\end{align}
with $\bm{\hat{\theta}}^{\bm{u}}\left(\cdot\right)$ a generally nonlinear operator and 
$\ptune\left(k\right)$ the most recent element of $\Ktune_s$ before control time step $k$. 
The updated type-2 fuzzy sets in the antecedents of the fuzzy rules of the fuzzy 
controller are given by
\begin{align}
\Big\{
&X_{1,i,s,q}:X_{\nx_{i,s},i,s,q}, \ q\in\left\{\pi_{\delta}(k),\ldots,\pz{k},k\right\}; \\
&N_{1,i,s,q}:N_{\nx_{i,s},i,s,q}, \ q\in\left\{\pi_{\delta}(k),\ldots,\pz{k},k\right\}\Big\}
=\hat{\pi}^{\bm{u}}\left(\tuant_{i,s}\left(k\right)\right),\nonumber
\end{align}   
where $\hat{\pi}^{\bm{u}}(\cdot)$ is a generally nonlinear operator.%

The global control objective is to reduce the realized cumulative value 
$\sum_{s=1}^{\ns}\hat{J}_s\left(\xe_s\left(l\right),
\bm{u}_s\left(l\right), \bm{u}_s\left(l - 1\right) \right)$ of a predefined cost function 
$\hat{J}_s\left(\cdot\right)$ for all subsystems $s$, $s\in\left\{1,\ldots,\ns\right\}$, 
by the end of the control procedure.    
Correspondingly, in designing the tuning procedure for $\tucon_{i,s}$, reduction of the cumulative 
cost will be taken into account. 
Additionally, the mutual interactions of the fuzzy-control agents that may influence 
the performance and local costs of the dynamically connected subsystems should be considered. 
Therefore, supposing that the number of control inputs of 
subsystem $s$ is $\Nu_s$,  
the MPC module solves the following minimization problem within the time window 
$\Ltune_s\left(k\right)$ to tune the consequent parameters of the control fuzzy rules, 
where $\bar{\bm{\theta}}^{{\bm{u},{\rm con}}}_s \left(k\right)
:= \left[\tucon_{1,s}(k):\tucon_{\Nu_s,s}(k)\right]^\top$ is the optimization variable:   
\begin{align}
\label{eq:consequent_optimizing_module}
&\min_{\bar{\bm{\theta}}^{{\bm{u},{\rm con}}}_s \left(k\right)}\hspace*{-.5ex}
\bigg(\hspace*{-.5ex}
w^{\rm p}\cdot\mathsmaller{\sum}\limits_{l \in \Ltune_s(k)} 
\hat{J}_s\Big(\xe_s(l), 
\bm{u}_s\left(l\right),
\bm{u}_s(l - 1 )\Big)+\nonumber\\
&\hspace*{8ex} w^{\rm f}. 
\Big|\phi_s(k)-\hspace*{-2ex} \mathsmaller{\sum}\limits_{l=k}^{k + \np - 1} 
\hat{J}_s\Big(\xe_s\left(l\right), 
\bm{u}_s\left(l\right)
,\bm{u}_s\left(l - 1 \right)\Big)
\Big|\bigg),\\
&{\rm s.t.} 
\left\{
\begin{array}{l}
\hspace*{-1ex}\text{\eqref{eq:subsystem_fuzzy_model} and \eqref{eq:agent_fuzzy_controller}}\\
\hspace*{-1ex}\bar{\bm{\theta}}^{{\bm{u},{\rm con}}}_s \left(l\right) = 
\bar{\bm{\theta}}^{{\bm{u},{\rm con}}}_s \left(k\right)
\end{array}
\right. 
\text{for $l\in\Ltune_s\left(k\right)\cup\left\{k,\ldots, k+ \np - 1\right\}$},\nonumber
\end{align}
with $w^{\rm p}$ and $w^{\rm f}$ the weighting factors for, respectively, the past and the future 
cumulative cost values, $\phi_s(k)$ an optimal value for the future cumulative local cost 
of subsystem $s$ that is computed by the MPC module (details on the computation of $\phi_s(k)$ will 
be given in Section~\ref{sec:MPC_controller}), and $\np$ the prediction horizon of the MPC module.  
Note that for computation of the past and future cumulative cost
values, the fuzzy model and the fuzzy-control agent should be run in a
loop (see the top control layer in
Figure~\ref{fig:two_layer_controller}) in order to produce the
estimated states and control inputs, assuming that the condition
$\bar{\bm{\theta}}^{{\bm{u},{\rm con}}}_s \left(l\right) =
\bar{\bm{\theta}}^{{\bm{u},{\rm con}}}_s \left(k\right)$ holds.  In
Figure~\ref{fig:two_layer_controller}, the past and future values of
the cost for subsystem $s$ are indicated by $\hat{J}^{\rm p}_s$ and
$\hat{J}^{\rm f}_s$, respectively.  The optimization problem
\eqref{eq:consequent_optimizing_module} is in general a nonlinear,
nonsmooth, and nonconvex problem, and can be solved by standard
optimization algorithms.%
\smallskip

\begin{remark}
The MPC module provides coordination among the distributed  fuzzy-control agents 
by optimizing the global cumulative cost of the entire system and computation of 
the corresponding optimal values of the local costs $\phi_s(k)$ for the  dynamically connected 
subsystems.
\end{remark}

\section{Top layer: MPC module}
\label{sec:MPC_controller}

\begin{figure}
\begin{center}
\psfrag{x1}[][][.8]{$\bm{x}_1(k)$}
\psfrag{x12}[][][.8]{$\xcor{12}(k)$}
\psfrag{u1}[][][.8]{$\bm{u}_1(k)$}
\psfrag{x2}[][][.8]{$\bm{x}_2(k)$}
\psfrag{x21}[][][.8]{$\xcor{21}(k)$}
\psfrag{u2}[][][.8]{$\bm{u}_2(k)$}
\psfrag{v1}[][][.8]{$\bm{d}_1(k)$}
\psfrag{v2}[][][.8]{$\bm{d}_2(k)$}
\psfrag{system}[][][.9]{{\color {blue}System}}
\psfrag{A1}[][][.7]{Fuzzy-control agent~1}
\psfrag{A2}[][][.7]{Fuzzy-control agent~2}
\psfrag{subsystem 1}[][][.7]{Subsystem~1}
\psfrag{subsystem 2}[][][.7]{Subsystem~2}
\includegraphics[width=.86\linewidth]{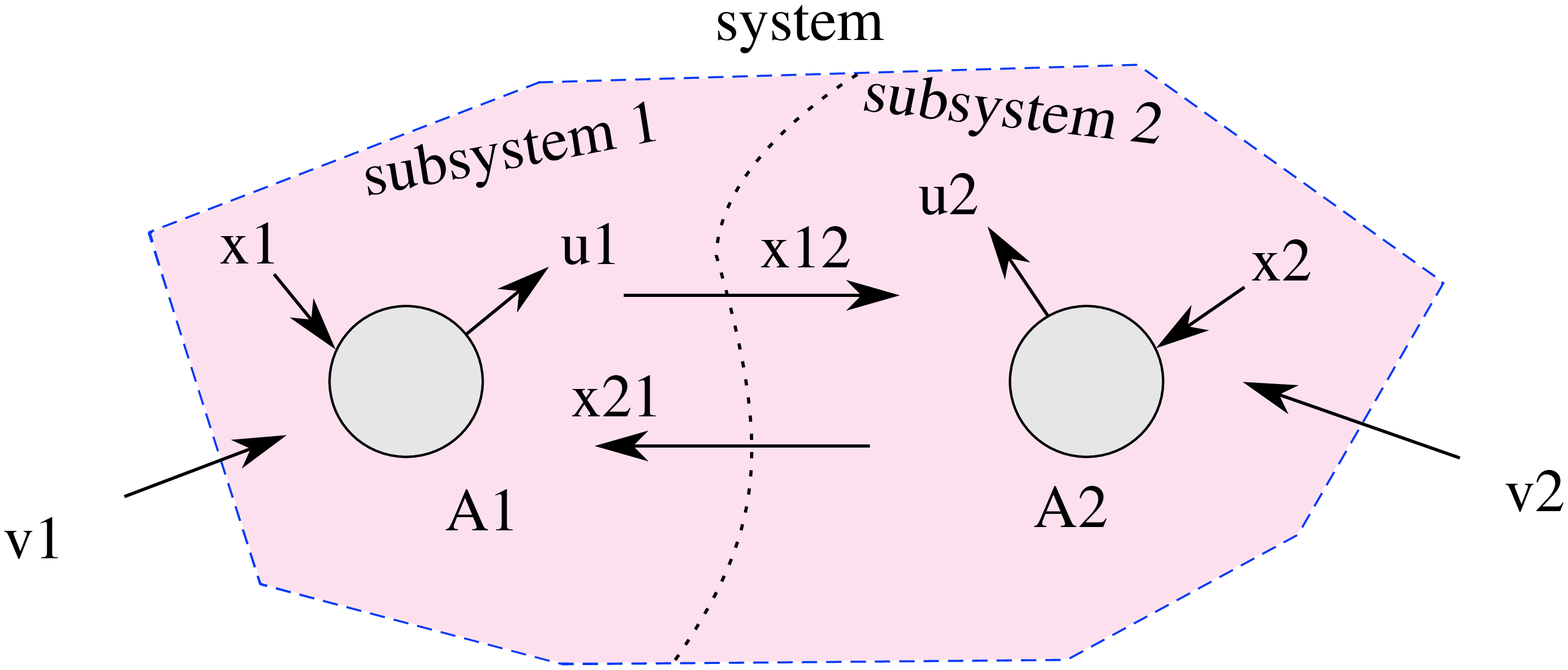}
\caption{Two dynamically interconnected subsystems.}
\label{fig:connected_subsystems}
\vspace*{2ex}
%
\psfrag{phi1}[][][.8]{$\phi_1(k)$}
\psfrag{phi2}[][][.8]{$\phi_2(k)$}
\psfrag{x12}[][][.8]{$\bm{x}_{12}(k)$}
\psfrag{x21}[][][.8]{$\bm{x}_{21}(k)$}
\psfrag{xe}[][][.8]{$\bm{x}(k)$}
\psfrag{u}[][][.8]{$\bm{u}(k)$}
\psfrag{d1}[][][.8]{$\bm{d}_1(k)$}
\psfrag{d2}[][][.8]{$\bm{d}_2(k)$}
\psfrag{system}[][][.8]{{\color {blue}System}}
\psfrag{fuzzy controller1}[][][.8]{Fuzzy-control agent~1}
\psfrag{model1}[][][.8]{Fuzzy model~1}
\psfrag{fuzzy controller2}[][][.8]{Fuzzy-control agent~2}
\psfrag{model2}[][][.8]{Fuzzy model~2}
\psfrag{system1}[][][.8]{Subsystem~1}
\psfrag{system2}[][][.8]{Subsystem~2}
\psfrag{1st}[][][.8]{{\color{red}Bottom layer}}
\psfrag{2nd}[][][.8]{{\color{red}Top layer}}
\psfrag{optimize}[][][.8]{Optimizer \eqref{eq:consequent_optimizing_module}}
\psfrag{mpc}[][][.8]{MPC module}
\psfrag{model}[][][.8]{Integrated fuzzy model}
\includegraphics[width=\linewidth]{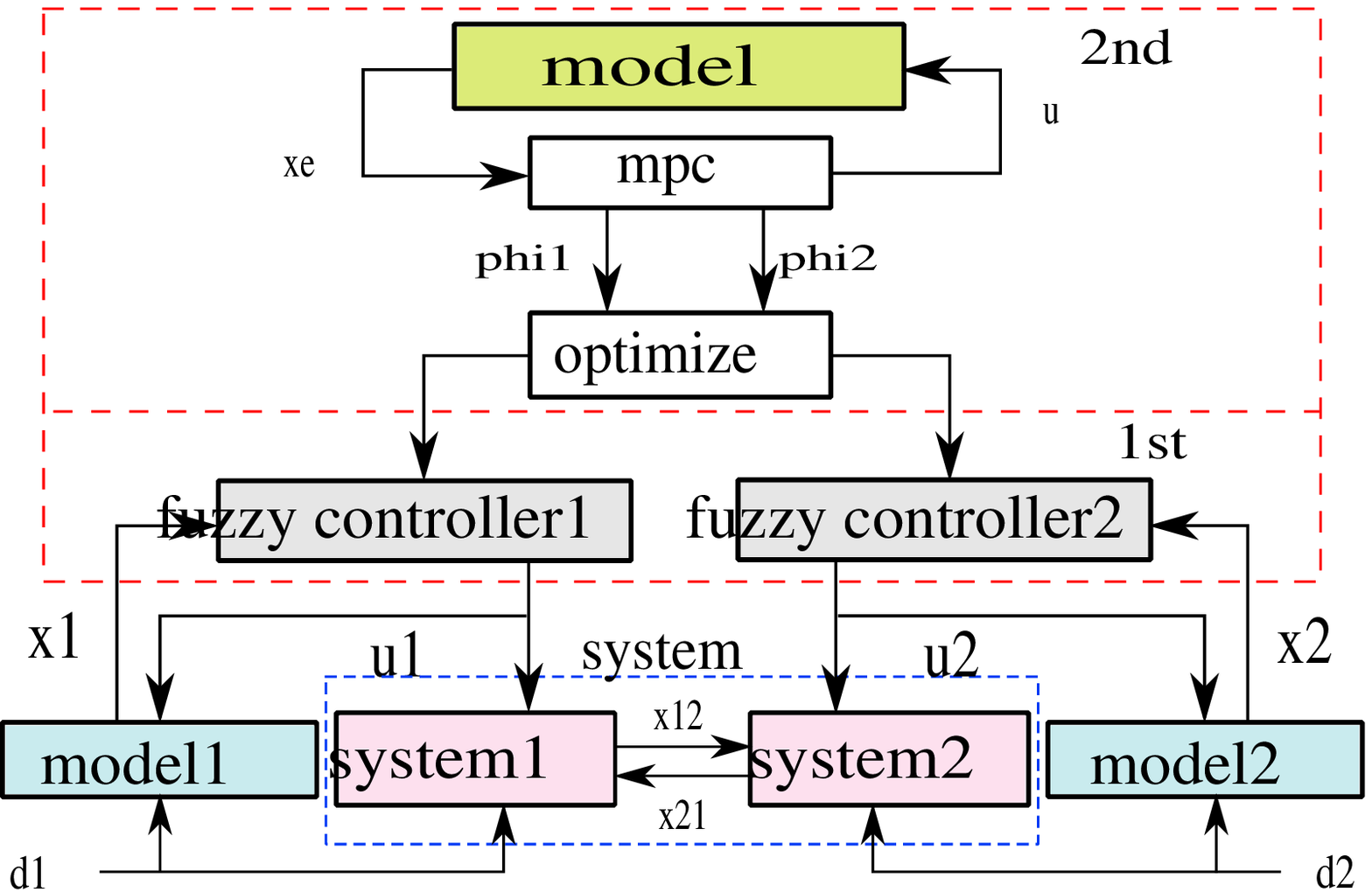}
\caption{Block diagram of the proposed two-layer control system  
for two dynamically interconnected subsystems.}
\label{fig:mpc_module}
\end{center}
\end{figure}

In this section, we explain the performance of the MPC module 
in more detail.  For the sake of simplicity and brevity, we give the
formulas for two dynamically connected subsystems~1 and 2 (see
Figure~\ref{fig:connected_subsystems}).
We assume that the fuzzy-control agents of the two subsystems have synchronized control time steps. 
The state and control input vectors of subsystem $s$, $s\in\left\{1,2,\right\}$,  
at control time step $k$ are denoted by $\bm{x}_s(k)$ and $\bm{u}_s(k)$. 
The vectors $\bm{x}(k)$ and $\bm{u}(k)$ are the state and control input 
vectors of the entire system (i.e., $\bm{x}(k) = \left[\bm{x}_1^\top(k), \bm{x}_2^\top(k) \right]^\top$ 
and $\bm{u}(k) = \left[\bm{u}_1^\top(k), \bm{u}_2^\top(k) \right]^\top$). 
Note that to keep the notations simple, in this section we avoid distinguishing the measured 
and estimated state variables. 
The external (both to the subsystem and to the entire system) disturbances 
that affect subsystem $s$ at control time step $k$ are indicated by $\bm{d}_s(k)$.   
The elements of the state vector of subsystem~1 that directly affect the dynamics of 
subsystem~2 are kept in a single vector $\xcor{12}(k)$ at control time step $k$, which is a 
subvector of $\xcor{1}(k)$. 
Subvector $\xcor{21}(k)$ of $\xcor{2}(k)$ is defined similarly. We have
\begin{equation}
\begin{aligned}
\label{eq:coordinated_noises}
&\vcor{1}(k) = \left[\bm{d}_1^\top\left(k\right) , \xcor{21}^\top(k)\right]^\top,\\
&\vcor{2}(k) = \left[ \bm{d}_2^\top\left(k\right) , \xcor{12}^\top(k)\right]^\top,
\end{aligned}
\end{equation}
where $\vcor{1}(k)$ and $\vcor{2}(k)$ are the total external 
disturbances for subsystems~1 and 2 at time step $k$  
and $\xcor{21}(k)$ and $\xcor{12}(k)$ are the external disturbances 
(only for the subsystems and not for the entire system) that affect, 
respectively, subsystem~1 and 2.%

Figure~\ref{fig:mpc_module} shows the block diagram of the proposed two-layer control 
system  applied  to these two subsystems. 
In this case, we consider a prediction horizon of 3 for the MPC module, 
for the following reasons. 
The control input $\bm{u}_s\left(k\right)$ of subsystem $s$ at control time step $k$,  
affects the cumulative cost of the entire system both at the current and upcoming 
control time steps $k$  and $k+1$ 
(see the formulation of $\hat{J}\left(\cdot\right)$ in \eqref{eq:average_step_cost}). 
Moreover, the effect of the control input $\bm{u}_s\left(k\right)$ of
subsystem $s$ at control time step $k$, is observed on the state of
the subsystem at control time step $k+1$, when
$\bm{x}_s\left(k+1\right)$ (or a subvector of it) will also act as
external disturbance for the other subsystem (see
\eqref{eq:coordinated_noises}).  The influence of this disturbance on
the state variable of the other subsystem will only be observed at the
next control time step, $k+2$.  Since $\bm{u}_s\left(k\right)$ will
affect the cumulative cost of the entire system at control time steps
$k$, $k+1$, and $k+2$, 
the minimum required size for the MPC prediction horizon is 3.

The centralized optimization problem  that should be solved by the MPC module in the 
top control layer (see Figure~\ref{fig:mpc_module}) to determine the optimal control 
inputs for both subsystems, such that the global cumulative cost of the entire system is 
minimized, for a prediction horizon of 3 is formulated by
\begin{align}
\label{eq:centralized_MPC}
& \hspace*{7ex} \min_{\tilde{\bm{u}}\left(k\right)}\quad \hat{\pazocal{J}}\left(k,\np\right)
\mathsmaller{\equiv}\\
&\min_{ \bm{u}\left(k\right) , \tilde{\bm{u}}_1\left(k + 1\right), \tilde{\bm{u}}_2\left(k + 1\right)}
\hspace*{.5ex}\mathsmaller{\sum}\limits_{s=1}^2 \mathsmaller{\sum}\limits_{l=k}^{k+2} 
\hat{J}_s\left(\bm{x}_s\left(l\right), \bm{u}_s\left(l\right), \bm{u}_s\left(l - 1\right)\right), \nonumber\\
&\rm{s.t.} 
\left\{
\begin{array}{ll}
\ueq\left(\tilde{\bm{u}}(l) \right) = 0 & \\
\uneq\left(\tilde{\bm{u}}(l) \right) \geq 0 & \text{ for } l\in\left\{ k,k+1,k+2 \right\}\\
\text{integrated }\eqref{eq:subsystem_fuzzy_model}\text{ for } s\in\{1,2\}, 
\end{array}
\right.\nonumber
\end{align}
with $\hat{\pazocal{J}}\left(k,\np\right)$ the global cumulative cost of the entire system  
within the MPC prediction window, 
$\tilde{{\bm{u}}}(k)$ a vector that includes all vectors ${\bm{u}}$ within the MPC prediction window, 
$\tilde{\bm{u}}_s(k + 1) = \left[\bm{u}_{s}^\top(k + 1), \bm{u}_s^\top(k + 2) \right]^\top$ 
for $s\in\{1,2\}$, 
and $\ueq\left(\cdot\right)$ and $\uneq\left(\cdot\right)$ operators that give the 
equality and inequality constraints for the control inputs of the subsystems.    
An \emph{integrated} fuzzy model (see Figure~\ref{fig:mpc_module}) should be 
used for the dynamics of the entire system, which implies that for each subsystem at every control 
time step, the terms $\xcor{12}$ and $\xcor{21}$ will affect the disturbances $\bm{\nu}_1$ and $\bm{\nu}_2$ 
(see \eqref{eq:coordinated_noises}) applied to \eqref{eq:subsystem_fuzzy_model}.%

Note that in \eqref{eq:subsystem_fuzzy_model}, we can define $\fx_{i,s}(\cdot)$ as a convex function. 
Based on Remark~\ref{remrak:combined_fuzzy_model}, the resulting fuzzy model for each subsystem 
will be both convex and smooth. 
Supposing that $\hat{J}_s(\cdot)$,  $\ueq\left(\cdot\right)$,  and $\uneq\left(\cdot\right)$ 
are also convex and smooth, then the optimization problem of \eqref{eq:centralized_MPC} will be 
a convex optimization problem that can be solved efficiently by gradient-based methods.  
The MPC module should compute and send the optimal values, $\phi_s(k) = 
\mathsmaller{\sum}\limits_{l=k}^{k+2} 
\hat{J}_s\left(\bm{x}_s\left(l\right), \bm{u}_s\left(l\right), \bm{u}_s\left(l - 1\right)\right)$, 
of the local costs to the subsystems (see Figures~\ref{fig:two_layer_controller} and \ref{fig:mpc_module}). 
The optimization problem \eqref{eq:centralized_MPC} has a decomposable structure \cite{Palomar:2006}, 
and in case the size of the problem increases or the centralized solution becomes too complex or 
costly to determine, it can be solved via the primal decomposition method \cite{Bemporad:2010} with 
$\tilde{\bm{u}}_1\left(k + 1\right)$ and $\tilde{\bm{u}}_2\left(k + 1\right)$ 
the private and $\bm{u}\left(k\right)$ the complicating variables.%

\section{Case study: Traffic modeling and control}
\label{sec:case_study_and_results}

To assess the proposed modeling and control approaches, we next perform a case study 
for an urban traffic network.%

\subsection{Setup}
\label{sec:setup}


\begin{figure}
\begin{center}
\psfrag{R}[][][.6]{R}
\psfrag{L}[][][.6]{L}
\psfrag{1R}[][][.6]{1R}
\psfrag{2R}[][][.6]{2R}
\psfrag{3R}[][][.6]{3R}
\psfrag{4R}[][][.6]{4R}
\psfrag{5R}[][][.6]{5R}
\psfrag{6R}[][][.6]{6R}
\psfrag{7R}[][][.6]{7R}
\psfrag{1L}[][][.6]{1L}
\psfrag{2L}[][][.6]{2L}
\psfrag{3L}[][][.6]{3L}
\psfrag{4L}[][][.6]{4L}
\psfrag{5L}[][][.6]{5L}
\psfrag{6L}[][][.6]{6L}
\psfrag{7L}[][][.6]{7L}
\psfrag{N}[][][.6]{N}
\psfrag{S}[][][.6]{S}
\psfrag{W}[][][.6]{W}
\psfrag{E}[][][.6]{E}
\includegraphics[width=0.85\linewidth]{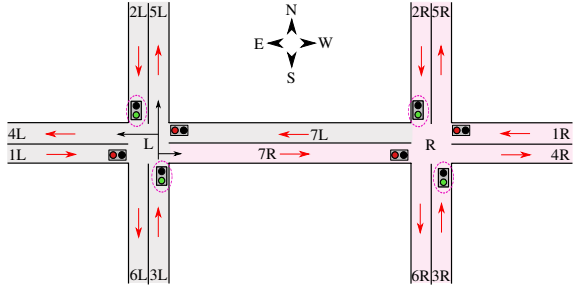}
\caption{Urban traffic network used for the case study.}
\label{fig:network}
\end{center}
\end{figure} %

The traffic network shown in Figure~\ref{fig:network} consists of two intersections, which have been indicated 
by ``L'' and ``R'' in the figure\footnote{
Note that the modeling and control approaches proposed in this paper are potentially suited 
for large-scale systems. In the given cased study, however, we have restricted ourselves to only 
two subsystems in order to provide more insight and an easier representation of the results.}. 
Each link of this urban traffic network consists of two lanes, which in Figure~\ref{fig:network} 
have been indicated by 1L, \ldots, 7L, and 1R, \ldots, 7R. 
For intersections L, lanes 1L, 2L, 3L, and 7L, and lanes 4L, 5L, 6L, and 7R  act as, respectively,  
the entrance and exit lanes (see the direction of the red arrows in Figure~\ref{fig:network}). 
Similarly, lanes 1R, 2R, 3R, and 7R are the entrance lanes for intersection R, while lanes 
4R, 5R, 6R, and 7L are the exit lanes for this intersection.%

In the rest of the paper, lanes with the indication numbers 1, \ldots, 6 are referred to as 
``side lanes'', and lanes with the indication number 7, as the ``connecting lanes''. 
Turning (see the black arrows in Figure~\ref{fig:network}) is allowed for the vehicles at intersections, 
except for U-turns.   
Every intersection has four traffic signals, each controlling all the rights-of-way of the entrance lane 
on which the traffic light stands. 
The traffic signals at the opposite entrance lanes of an intersection are synchronized and follow the same schedule 
(i.e., the green and red phases of the northern/southern traffic signals in Figure~\ref{fig:network}, 
as well as those of the western and eastern ones coincide). 
The length of the side and connecting lanes are 150~m and 300~m respectively, the average vehicle length 
(including the safety distances from the back and front vehicles) is 7.5~m, and the cycle time of the traffic 
signals for both intersections is 90~s.%

\subsection{Modeling}
\label{sec:case_study_modeling}

The urban traffic network 
is divided into two subnetworks, called ``subnetwork~1'' and
``subnetwork~2'', colored in, respectively, grey and pink in the
figure.  Subnetwork~1 consists of intersection L and lanes 1L, \ldots,
7L, and subnetwork~2 includes intersection R and lanes 1R, \ldots, 7R.
Three different classes of fuzzy models describing the behavior of
traffic, are developed for each subnetwork: ``class~1'', including
type-1 membership functions, ``class~2'', including
probabilistic-fuzzy membership functions, and ``class~3'', including
fuzzy-fuzzy membership functions.  These models will be formulated for
two state variables, the total number of vehicles per link ($n$) and
the number of vehicles in the queue on a link ($q$), and will consist
of fuzzy rules with a formulation following
\eqref{eq:subsystem_fuzzy_model}.%
 
We assume that the most reliable measurements of the state variables available at control time 
step $k$ from the traffic sensors, correspond to control time step $k-1$.
Each fuzzy rule $r$ is described by: 
``if $\bm{x}(k-1) \in X_r$ $\land$ $\bm{u}(k-1) \in U_r$ $\land$ $\bm{\nu}(k-1) \in N_r$, 
then $\bm{x}_r(k) = a^{\bm{x}}_{0,r} + a^{\bm{x}}_{1,r}  \bm{x}(k-1) + a^{\bm{x}}_{2,r} \bm{u}(k-1) 
+ a^{\bm{x}}_{3,r}\bm{\nu}(k-1)$.'' 
For the traffic scenarios we have considered in this case study, the range of variations of the parameters 
in the given urban traffic network is limited. 
Hence, the parameters of the fuzzy sets $X_r$, $U_r$, and $N_r$ in the antecedents are assumed fixed, 
and only the parameter vectors $\left[a^{\bm{x}}_{0,r},a^{\bm{x}}_{1,r},a^{\bm{x}}_{2,r},
a^{\bm{x}}_{3,r}\right]^\top$ of the consequents 
will be identified and updated.%

The control variable of each subnetwork is the green time of the northern and southern 
traffic signals (indicated by the red dashed curves in Figure~\ref{fig:network}).
These traffic signals are synchronized, and the green time of the other two traffic signals of each intersection 
(which are also synchronized), is the difference between the cycle time of the intersection and the  
control input. 
The flows of the vehicles that enter the urban traffic network via the source lanes (1L, 2L, 3L, 1R, 2R, 3R) 
are considered as the external inputs.  
The fuzzy sets $X_r$ and $N_r$ to which the state variables and the external inputs of the urban 
traffic network belong, will each be defined for the two qualitative terms ``low'' and ``high''. 
Additionally, the fuzzy set $U_r$ to which the control inputs belong, will be defined for the two 
qualitative terms ``short'' and ``long''.  
Next, we explain the three different fuzzy membership functions that are used for the three classes of fuzzy models.%

\smallskip

\subsubsection{Type-1 triangular membership function}

\begin{figure}
\begin{center}
\includegraphics[width = .45\textwidth]{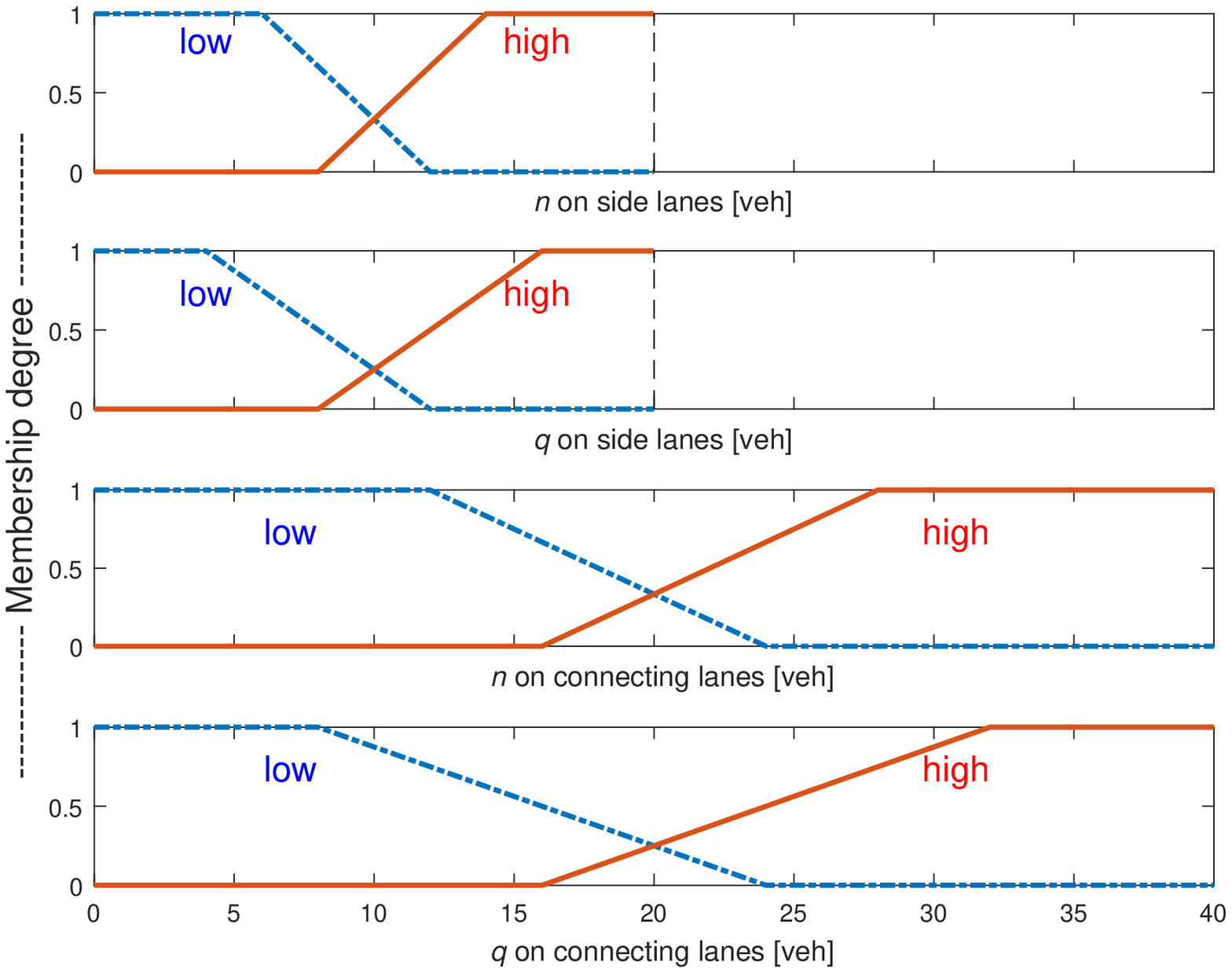}

\vspace*{3ex}

\includegraphics[width = .45\textwidth]{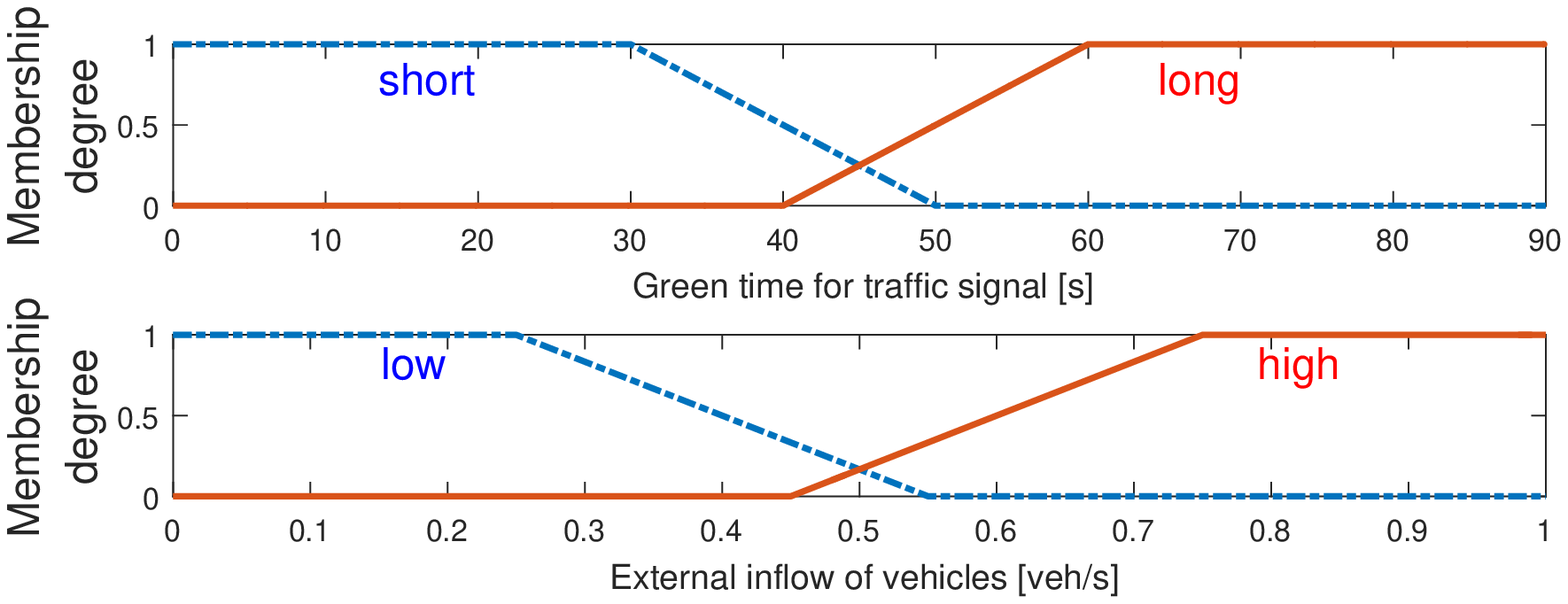}
\caption{Type-1 triangular membership functions for state variables, control inputs, and external 
inputs. \label{fig:triangular_MF}}
\end{center}
\end{figure}

For the fuzzy models in class~1, we consider type-1 triangular membership functions, 
for two main reasons: in addition to the simplicity and low computation time, 
Pedrycz \cite{Pedrycz:1994} shows that 
triangular membership functions with the half overlap level when used for modeling, 
can lead to entropy equalization. 
The type-1 triangular membership functions used for the models within class~1 are shown in Figure~\ref{fig:triangular_MF}.%
 
\smallskip

\subsubsection{Type-2 probabilistic-fuzzy membership function} 

\begin{figure}
\begin{center}
\includegraphics[width = 0.415\textwidth]{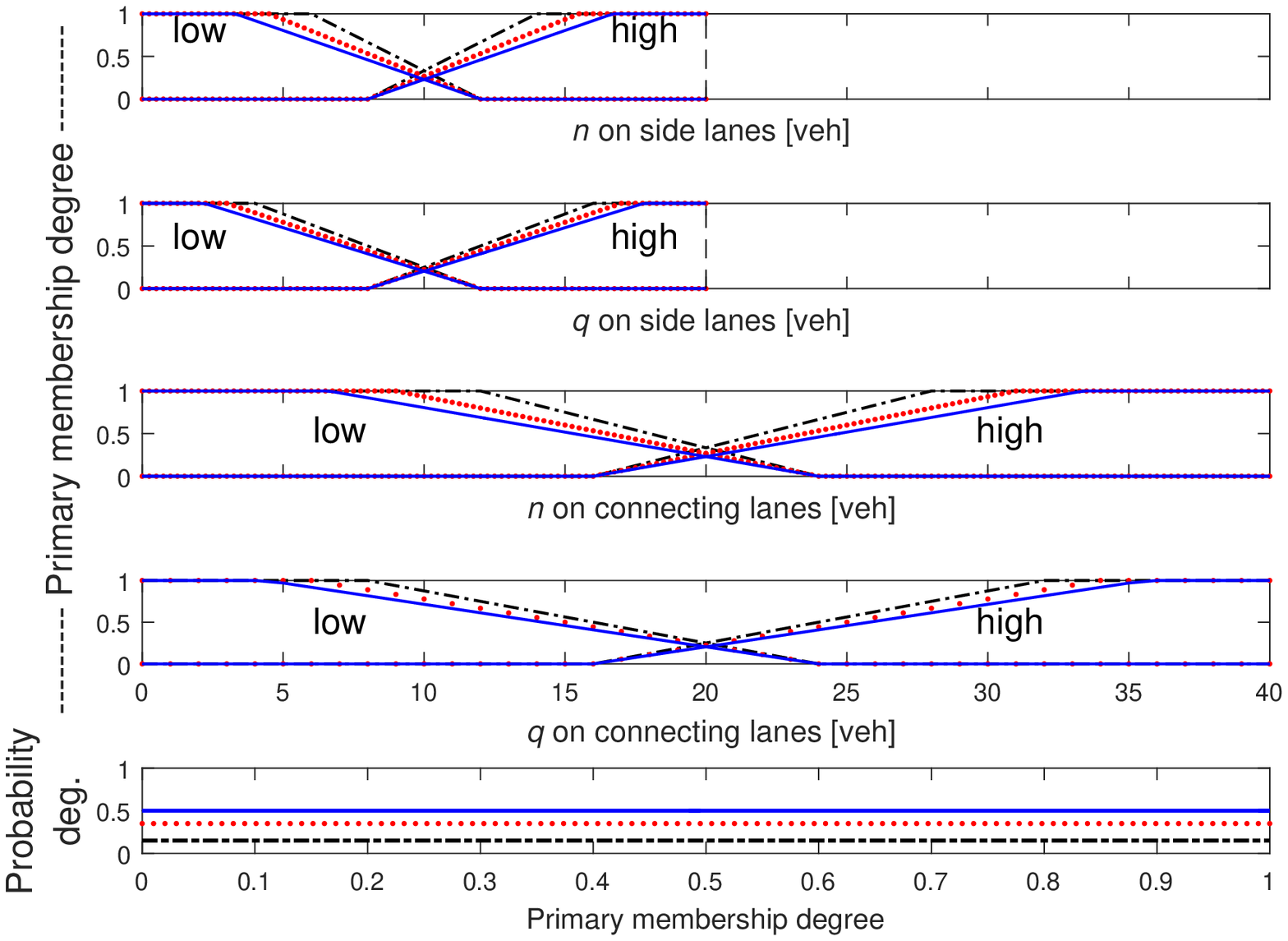}

\vspace*{3ex}

\includegraphics[width=0.415\textwidth]{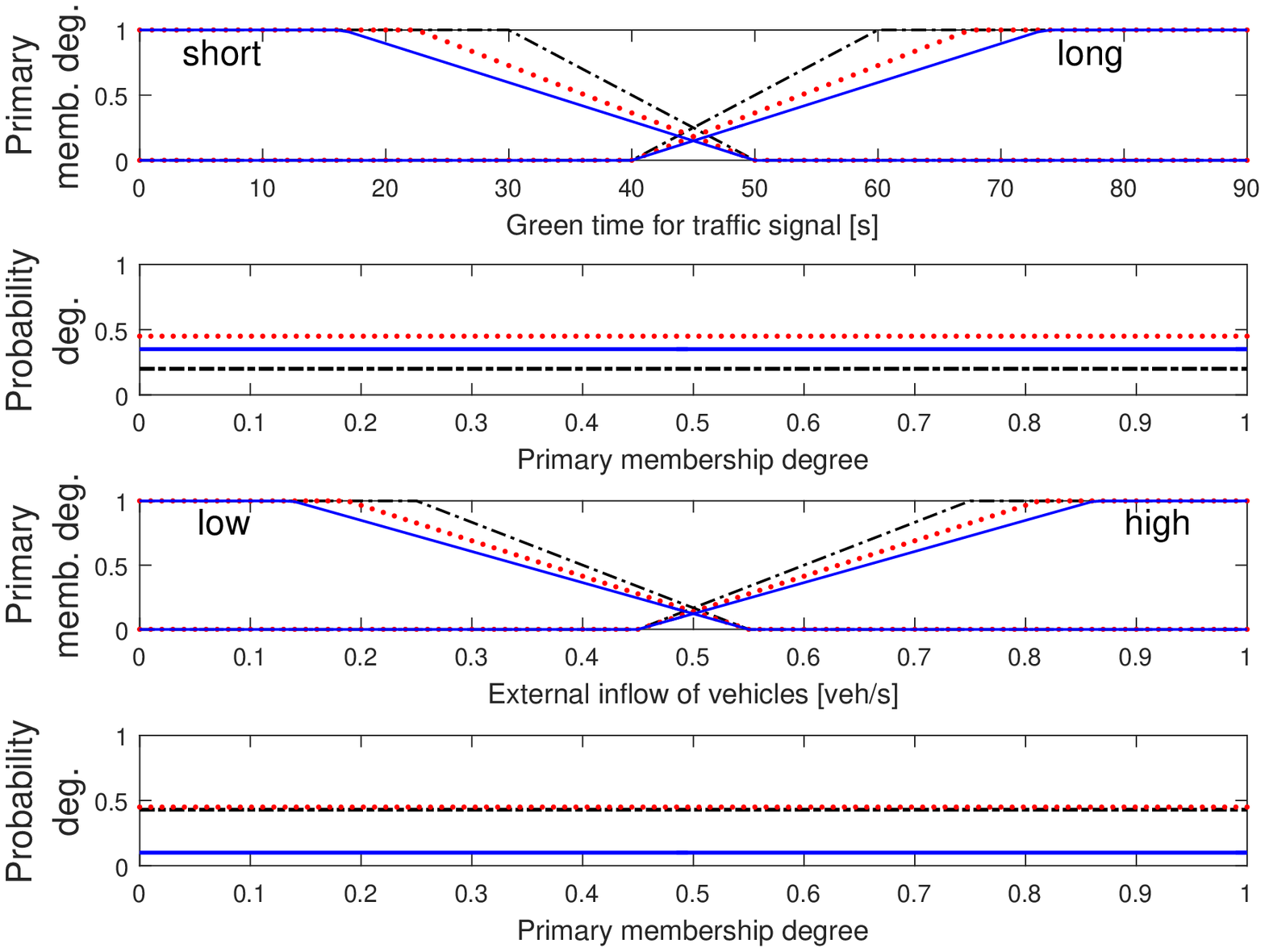}
\caption{Probabilistic-fuzzy membership functions for state variables, control inputs, and 
external inputs. \label{fig:ProbFuzzy_MF}}
\end{center}
\end{figure}

The primary and secondary membership functions of the type-2 probabilistic-fuzzy membership functions 
used in models within class~2, are illustrated in Figure~\ref{fig:ProbFuzzy_MF}. 
The secondary membership functions in this case are probability functions, which have been considered  
fixed-value. 
Each probability function and its corresponding primary membership function have been plotted using the same color. 
The antecedent of each rule includes three events defined on the state variable $\bm{x}$, the control input 
$\bm{u}$, and the external inputs $\bm{\nu}$.%

Each status \emph{low} and \emph{high}, or \emph{short} and \emph{long}, may be interpreted in three different ways 
(see the blue, red, and black plots in Figure~\ref{fig:ProbFuzzy_MF}). 
Suppose that $\event{x,\low}{1}$, $\event{x,\low}{2}$, and $\event{x,\low}{3}$ indicate the event  
``$\bm{x}$ is low'' for each of these three interpretations. Similarly, the events ``$\bm{x}$ is high'', 
``$\bm{u}$ is short'', ``$\bm{u}$ is long'', ``$\bm{\nu}$ is low'', and ``$\bm{\nu}$ is high'' 
for the three interpretations are indicated by, 
respectively, $\event{x,\high}{1}$,  $\event{x,\high}{2}$,  $\event{x,\high}{3}$, and 
 $\event{u,\short}{1}$, $\event{u,\short}{2}$, $\event{u,\short}{3}$, and 
 $\event{u,\lng}{1}$, $\event{u,\lng}{2}$, $\event{u,\lng}{3}$, 
 and $\event{\nu,\low}{1}$, $\event{\nu,\low}{2}$, $\event{\nu,\low}{3}$, 
 and $\event{\nu,\high}{1}$, $\event{\nu,\high}{2}$, $\event{\nu,\high}{3}$.%

Consider the antecedent of a specific rule that is described by ``if $\bm{x}$ is low 
 and $\bm{u}$ is long and $\bm{\nu}$ is high''; the three events $\event{x,\low}{i}$, 
 $\event{u,\lng}{j}$, and $\event{\nu,\high}{k}$ (for $i,j,k \in \{1,2,3\}$) should occur 
 at the same time, for this specific rule to be fired. 
The probability of occurrence of these three events simultaneously, 
 and hence, the activation of this specific fuzzy rule is given by
\begin{align}
\label{eq:prob_three_events}
\hat{p}&\left(\event{x,\low}{i} \land \event{u,\lng}{j} \land \event{\nu,\high}{k} \right)  = \\
&
\hat{p}\left(\event{x,\low}{i}\right)\cdot
\hat{p}\left(\event{u,\lng}{j} | \event{x,\low}{i} \right)\cdot 
\hat{p}\left(\event{\nu,\high}{k} | \event{x,\low}{i} \land \event{u,\lng}{j}\right), \nonumber
\end{align}
with $\hat{p}(\cdot)$ the probability function. 
For the sake of simplicity, we assume that the three events are independent, which 
results in the following simplified expression:
\begin{align}
\label{eq:prob_independent_events}
\hat{p}\left(\event{x,\low}{i} \land \event{u,\lng}{j} \land \event{\nu,\high}{k} \right)  = 
\hat{p}\left(\event{x,\low}{i}\right)\cdot
\hat{p}\left(\event{u,\lng}{j}\right)\cdot 
\hat{p}\left(\event{\nu,\high}{k}\right). 
\end{align} 
Therefore, the primary membership degree of the combined event  $\event{x,\low}{i} \land 
\event{u,\lng}{j} \land \event{\nu,\high}{k}$ 
is obtained by aggregation of the corresponding primary type-1 membership functions, or equivalently 
by determining the minimum or multiplication of the primary membership degrees of the events $\event{x,\low}{i}$, 
$\event{u,\lng}{j}$, and $\event{\nu,\high}{k}$. 
The probability that this combined membership degree is realized is computed by 
\eqref{eq:prob_independent_events}.

\smallskip

\subsubsection{Type-2 fuzzy-fuzzy membership function}

\begin{figure}
\begin{center}
\includegraphics[width = 0.415\textwidth]{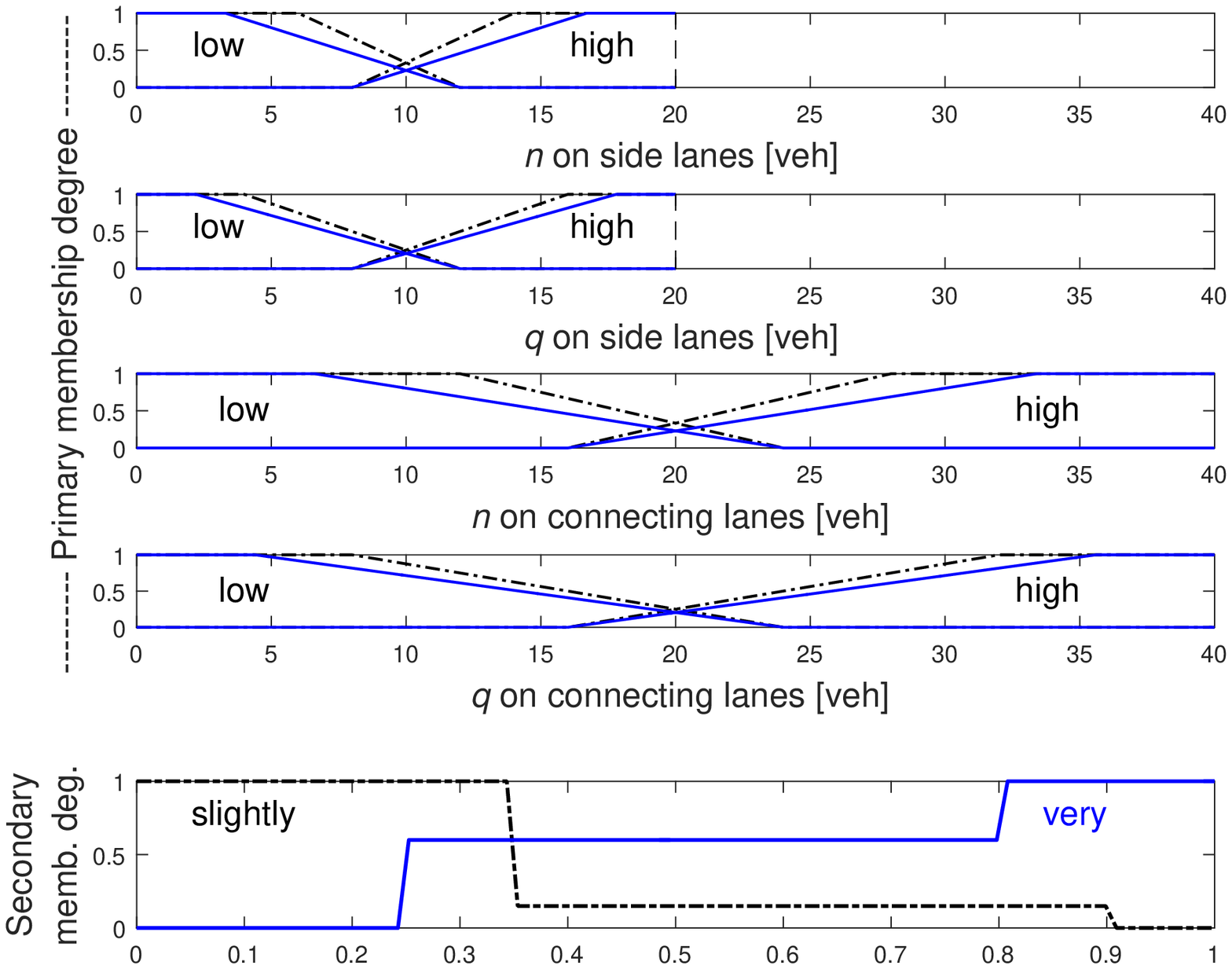}

\vspace*{4ex}

\includegraphics[width = 0.415\textwidth]{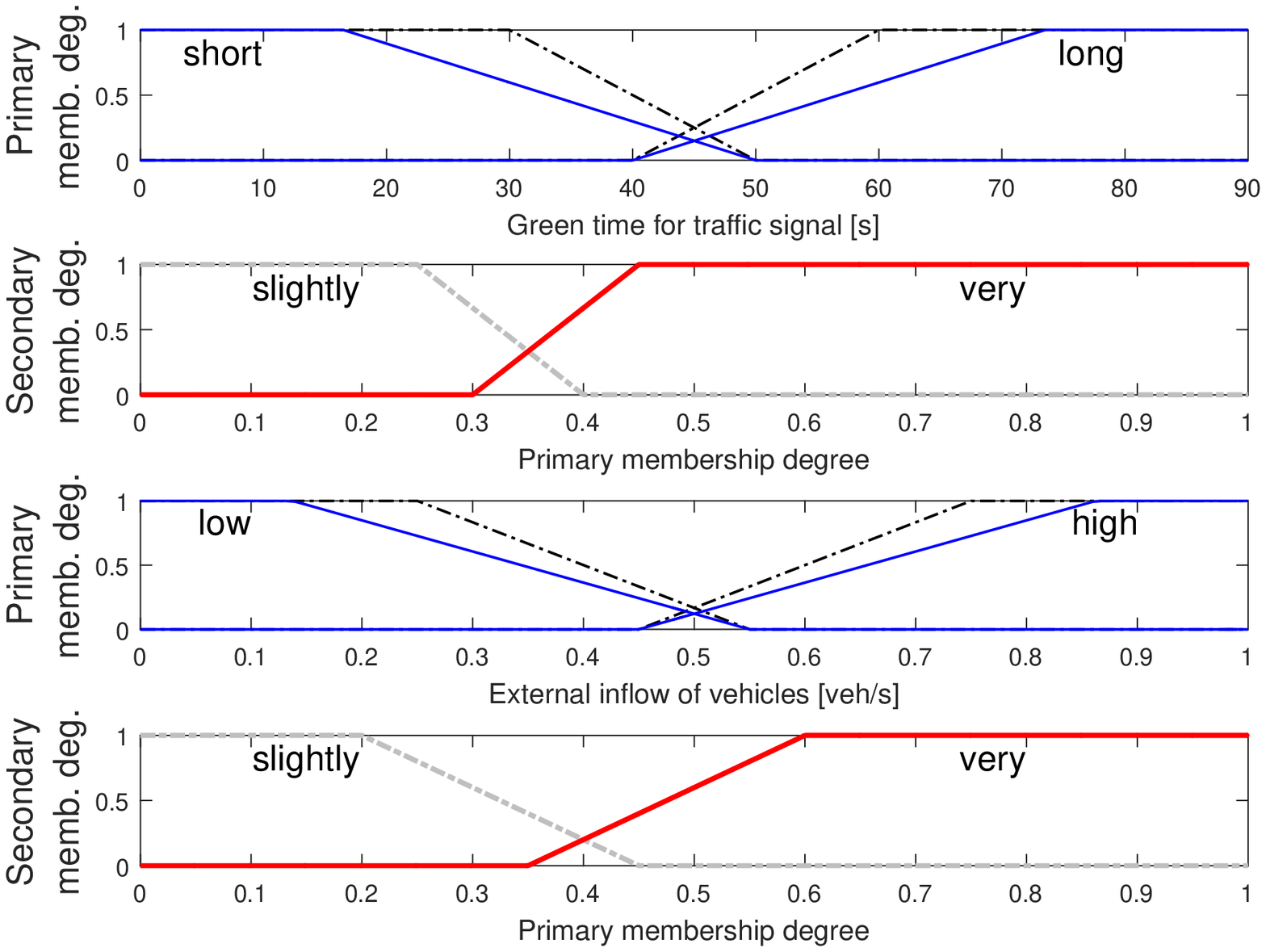}
\caption{Fuzzy-fuzzy membership functions for state variables, control inputs, and external inputs. 
\label{fig:FuzzyFuzzy_MF_ctrl_dist}}
\end{center}
\end{figure}

In order to develop a fuzzy model of the urban traffic network within class~3, we need to describe the 
status of the network by fuzzy-fuzzy events (see Section~\ref{sec:ProbFuzzy_FuzzyFuzzy_MF}). 
Therefore, in the linguistic description of the corresponding fuzzy rules, we consider a 
second qualitative term (``slightly'' or ``very'') for the descriptive terms \emph{low} and \emph{high}, 
and \emph{short} and \emph{long}. The mathematical representation of these qualitative terms, 
i.e., the type-1 primary and secondary membership functions, is illustrated in Figure~\ref{fig:FuzzyFuzzy_MF_ctrl_dist}.

For the type-2 probabilistic-fuzzy membership functions, we considered three different interpretations 
for each of the terms \emph{low} and \emph{high}, and \emph{short} and \emph{long} 
(see the black dashed-dotted, red dotted, and blue solid curves in Figure~\ref{fig:ProbFuzzy_MF}). 
For computing the output of the fuzzy-fuzzy membership functions, the corresponding 
fuzzy inference engine will make all the possible combinations of the membership degrees obtained 
from different interpretations for the state variable, the control input, and the external input. 
Therefore, considering higher numbers of interpretations for the fuzzy terms can result in 
a dramatic growth in the computational burden. 
This problem does not arise with probabilistic-fuzzy membership functions. 
To reduce the computational burden for the fuzzy-fuzzy membership functions 
and to make the computations feasible, 
we have therefore considered two interpretations instead of 
three for each of the terms \emph{low} and \emph{high}, and \emph{short} and \emph{long} 
(see the black dashed-dotted and blue solid curves in Figure~\ref{fig:FuzzyFuzzy_MF_ctrl_dist}).%

Consider the antecedent of a specific rule that is given by ``if
$\bm{x}$ is very high and $\bm{u}$ is very short and $\bm{\nu}$ is
slightly low''; the three fuzzy-fuzzy events involved in this
antecedent are indicated by $\event{x,{\rm V\, high}}{i}$,
$\event{u,{\rm V\, short}}{j}$, and $\event{\nu,{\rm S low}}{k}$ (for
$i,j,k\in\{1,2\}$), respectively, and their primary membership degrees
are given by $\mu^{x,{\rm V\, high}}_{1,i}$, $\mu^{u,{\rm V\,
      short}}_{1,j}$, and $\mu^{\nu,{\rm S\, low}}_{1,k}$. Similarly,
the secondary membership degrees corresponding to each of these
primary membership degrees are denoted by $\mu^{x,{\rm V\,
      high}}_{2,i}$, $\mu^{u,{\rm V\, short}}_{2,j}$, and
$\mu^{\nu,{\rm S\, low}}_{2,k}$.  To find the membership degree of the
combined fuzzy-fuzzy events for this antecedent, one should consider
all the possible combinations of the primary and secondary membership
degrees (in this case, 8 combinations are possible).
For each combination $c_{i,j,k}$, the primary and secondary membership
degrees of the combined event are computed by
\begin{align}
\mu^{\rm com}_{1,c_{i,j,k}} = \min\left\{
\mu^{x,{\rm V\, high}}_{1,i},\ 
\mu^{u,{\rm V\, short}}_{1,j},\ 
\mu^{\nu,{\rm S\, low}}_{1,k}
\right\},\\
\mu^{\rm com}_{2,c_{i,j,k}} = \min\left\{
\mu^{x,{\rm V\, high}}_{2,i},\     
\mu^{u,{\rm V\, short}}_{2,j},\      
\mu^{\nu,{\rm S\, low}}_{2,k}
\right\}.
\end{align}  
In case two combinations have the same primary membership degree, 
the one with the maximum secondary membership degree is kept, and the rest of the equal primary 
membership degrees and their corresponding secondary membership degrees will not be considered 
for computing the output of the fuzzy inference engine (for more details see \cite{Zimmermann:1996}).  
The output of the inference engine of the fuzzy rule is computed based on the approach used in \cite{Liang:1999}.%

The maximum number of fuzzy rules that can be constructed for a fuzzy model within class~3 
in this case is $4^3$ (i.e., 64), i.e., each of the statements in the antecedent of the fuzzy rules 
regarding the state variable and the external input can have four various descriptions within the set 
$\left\{ \text{slightly , very} \right\}  \times \left\{\text{low , high}\right\}$ and the statement 
regarding the control input adopts either of the 
four descriptions within the set $\left\{ \text{slightly , very} \right\} \times \left\{\text{short , long}\right\}$.   
This implies that the total number of parameters that should be identified for the type-2 
fuzzy-fuzzy model is $4(64)$, i.e., 256 parameters (note that each rule has 4 parameters in its consequent 
that should be identified). 
Recall that the number of rules in the rule base of the fuzzy models of class~2 was $2^3$ (i.e., 8), 
i.e., each of the statements in the antecedent of the fuzzy rules about the state variable and the 
external input could accept either of the two descriptions within the set $\left\{\text{low , high}\right\}$ 
and the statement about the control input could adopt either of the two descriptions within 
the set $\left\{\text{short , long}\right\}$.%

As explained before, the computational burden for class~3 of the fuzzy
models is very high, and may grow dramatically with the number of
fuzzy rules.  To reduce the computational burden of both the
identification procedure and the computations of the fuzzy inference
engine, we have considered 8 rules for class~3 of the fuzzy models,
just as for class~2.%
%

\smallskip

\subsubsection{Model identification}
\label{sec:case_study_model_identification}

\begin{figure*}
\begin{center}
\psfrag{Relative error}[][][.6]{Relative validation error (\%) for $n$, subnetwork~1}
\includegraphics[width = .35\textwidth]{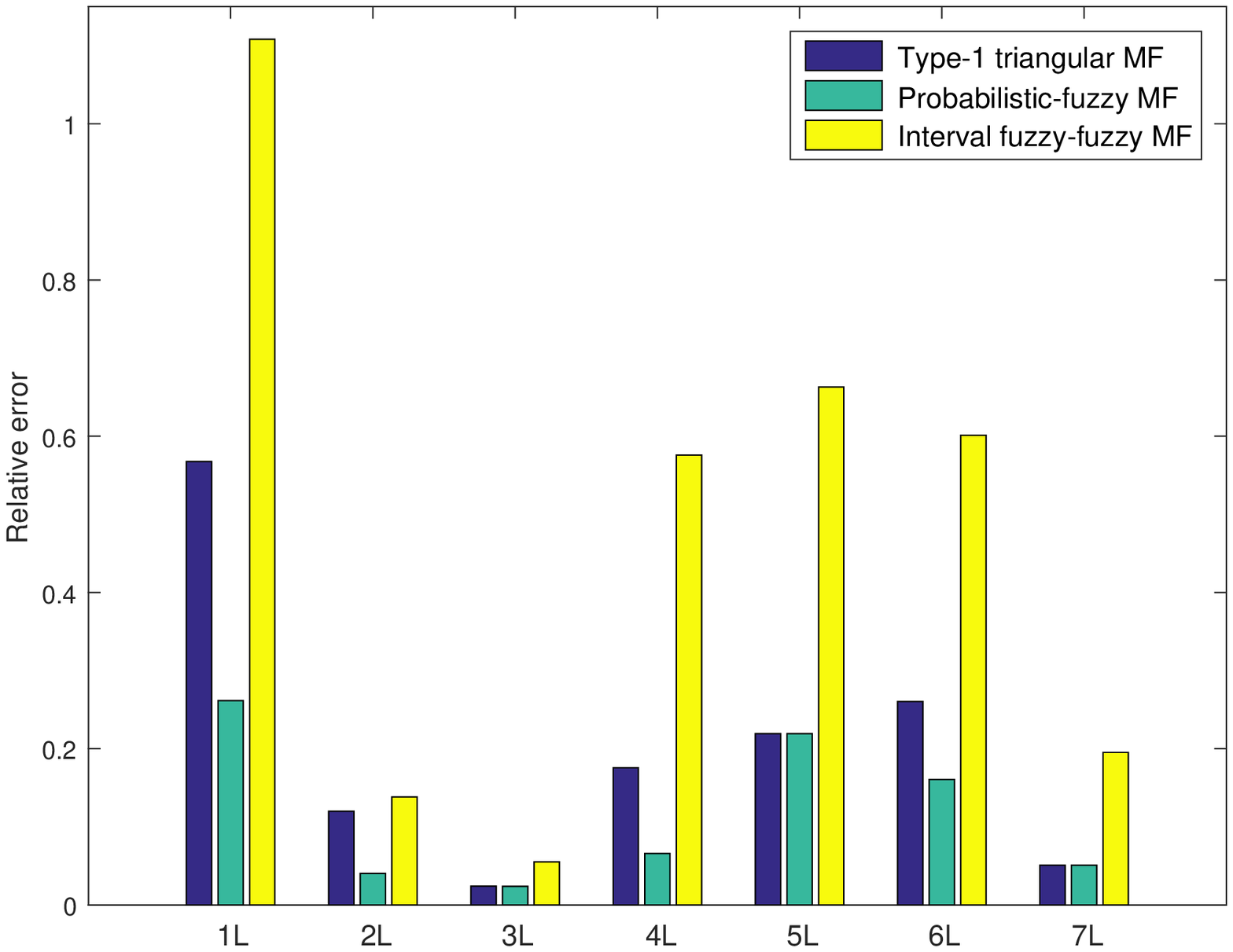}
\hspace*{3ex}
\psfrag{Relative error}[][][.6]{Relative validation error (\%) for $q$, subnetwork~1}
\includegraphics[width = .35\textwidth]{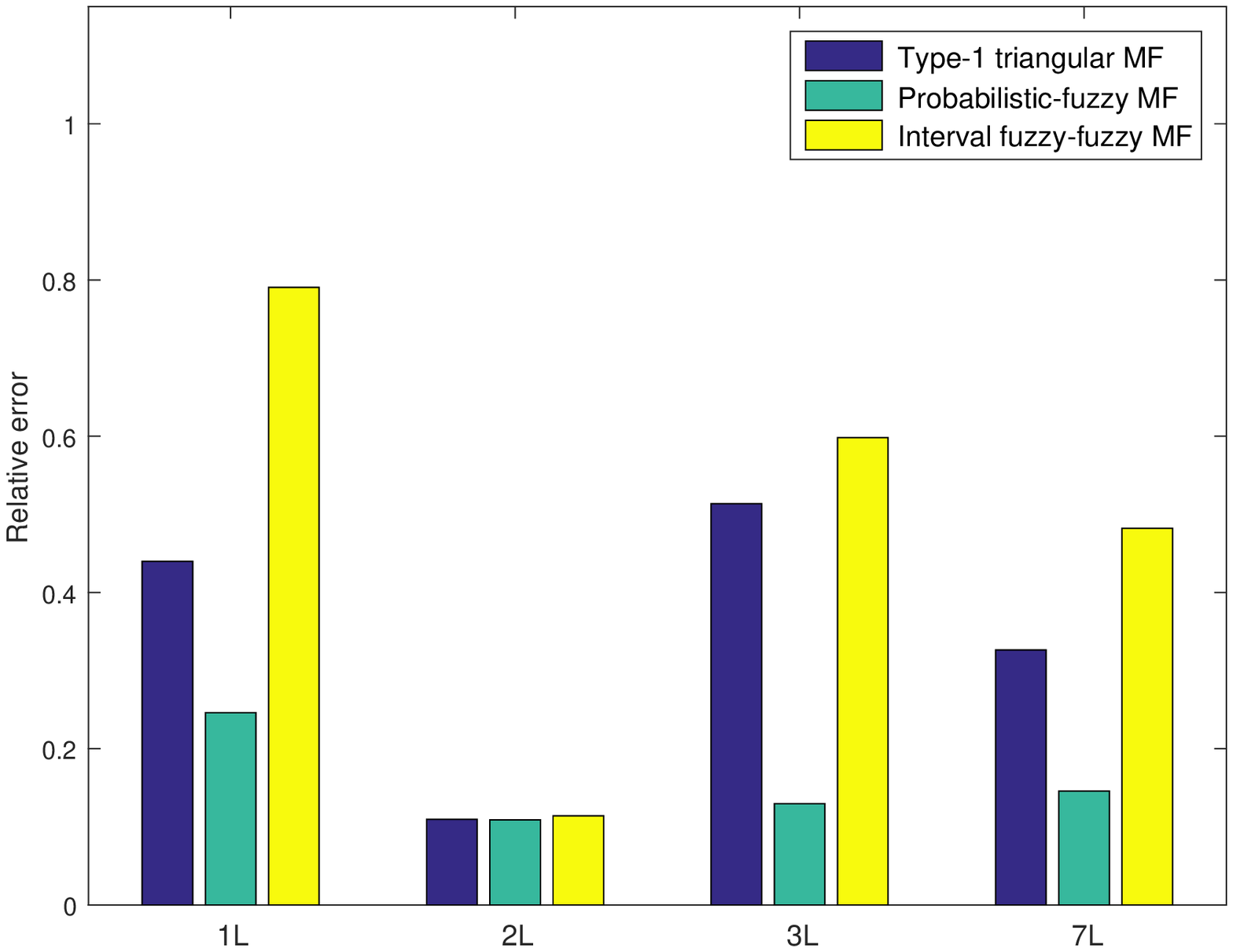}

\vspace*{3ex}

\psfrag{Relative error}[][][.6]{Relative validation error (\%) for $n$, subnetwork~2}
\includegraphics[width = .35\textwidth]{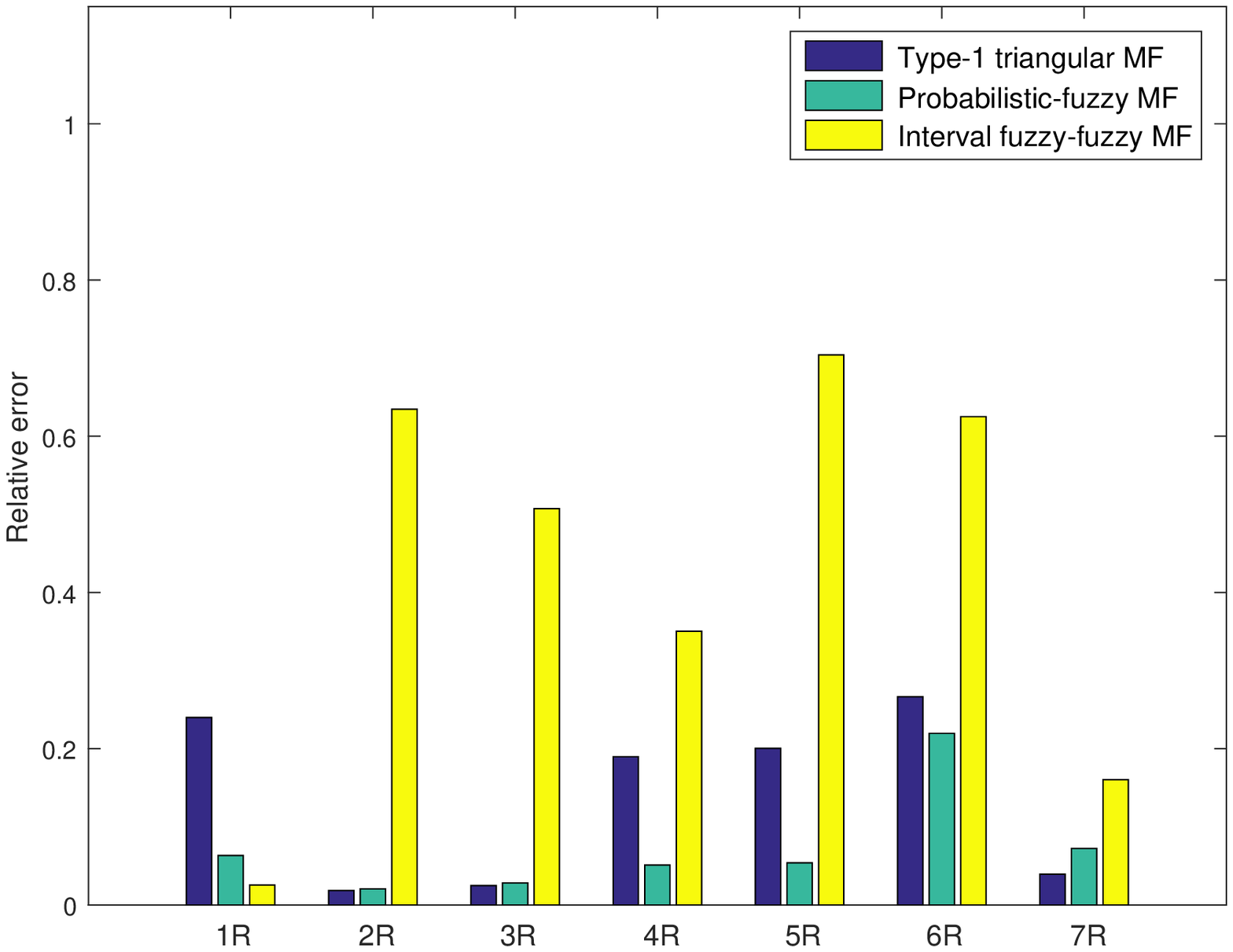}
\hspace*{3ex}
\psfrag{Relative error}[][][.6]{Relative validation error (\%) for $q$, subnetwork~2}
\includegraphics[width = .35\textwidth]{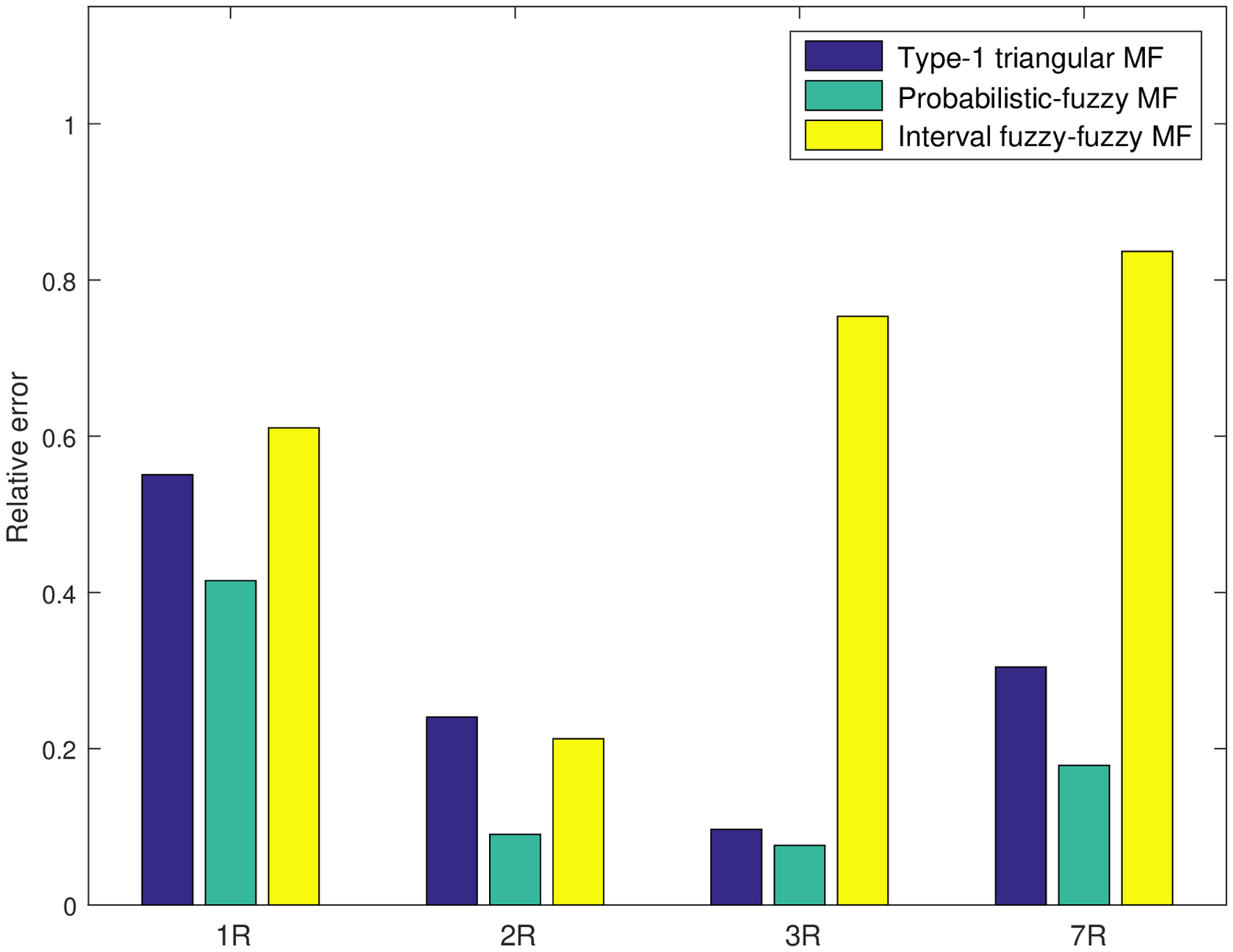}
\caption{Comparison of the relative validation errors (given as percentages) for estimation of the 
state variables  by the fuzzy models within different classes, for subnetwork~1 (plots in the first row) 
and subnetwork~2 (plots in the second row). 
\label{fig:relative_validation_errors}
}
\end{center}
\end{figure*}

Now, we explain the procedure of identifying the fuzzy models within class~1, class~2, and class~3. 
Data collected from the urban traffic network simulated within NetLogo \cite{NetLogo}, is used to 
identify and validate the fuzzy models. 
The urban traffic network illustrated in Figure~\ref{fig:network} was simulated in NetLogo, using 
Gipps' car following model \cite{Gipps:1981}. 
The data was collected in a 15~min run of the NetLogo simulator for random time-varying inflows of vehicles at 
the source lanes (these inflows have been considered such that all sources of the urban traffic network are  
exposed to low, moderate, and high inflows of vehicles). 
Data, including the state variables, the control inputs, and the external inputs were collected and saved, 
where 80\% of the collected data was used for identification and the other 20\% was used for validation of the 
fuzzy models.   
The relative validation errors of the three classes of fuzzy models
are shown in Figure~\ref{fig:relative_validation_errors}.  
Comparing the type-1 and the probabilistic-fuzzy model, we see that the latter has a lower 
validation relative error in almost all the cases. The fuzzy-fuzzy model is, however, the least 
accurate one among the three model classes. This can be due to the simplifications (reducing the number of 
primary membership functions from 3 to 2, and reducing the number of possible rules in the rule base) 
that had to be made to make the identification tractable, which in itself reveals a limitation for 
fuzzy-fuzzy membership functions.%

Note that the computation times for identifying the type-1 and the probabilistic-fuzzy models were very 
close to each other, while for the fuzzy-fuzzy model, this computation time was significantly higher. 
Therefore, the overall conclusion is that class~2 of the fuzzy models, which uses the newly proposed  
probabilistic-fuzzy membership functions, show clear advantage compared with class~1 and class~3 of 
models, which use type-1 and fuzzy-fuzzy membership functions.%

\subsection{Control}

In this section, two different control systems will be designed for the urban traffic network 
shown in Figure~\ref{fig:network}. 
The first control system includes a decentralized architecture, while in the second case, a 
coordinated control system is considered.    
In either of these two cases, each subnetwork will be controlled 
by one fuzzy controller that decides about the green time of the northern/southern traffic signal.%

The choice of a decentralized architecture for the comparison has two main reasons. 
First, since the computations are done by similar fuzzy controllers as those in the 
proposed integrated control architecture, both control systems will have almost similar computation times, 
which makes the comparison of the performances more fair. 
Second, with this comparison, we can see how significant the role of the second proposed 
layer including the MPC tuning module can be.%

The fuzzy rules of the fuzzy controllers corresponding to intersections L and R are defined by, respectively: \\
``if $\bar{\bm{x}}^{{\rm NS, L}}(k) \in X^{{\rm NS, L}}_r$ $\land$ $\bar{\bm{x}}^{{\rm EW, L}}(k) \in X^{{\rm EW, L}}_r$, 
then $\bm{u}^{\rm NS, L}_r(k) = a^{\bm{u},{\rm L}}_{0,r} + a^{\bm{u},{\rm L}}_{1,r} \cdot \bar{\bm{x}}^{{\rm NS, L}}(k) + 
a^{\bm{u},{\rm L}}_{2,r} \cdot \bar{\bm{x}}^{{\rm  EW, L}}(k)$'' 
and \\
``if $\bar{\bm{x}}^{{\rm NS, R}}(k) \in X^{{\rm NS, R}}_r$ $\land$ $\bar{\bm{x}}^{{\rm EW, R}}(k) \in X^{{\rm EW, R}}_r$, 
then $\bm{u}^{\rm NS, R}_r(k) = a^{\bm{u},{\rm R}}_{0,r} + a^{\bm{u},{\rm R}}_{1,r} \cdot \bar{\bm{x}}^{{\rm NS, R}}(k) + 
a^{\bm{u},{\rm R}}_{2,r} \cdot \bar{\bm{x}}^{{\rm  EW, R}}(k)$'', \\
with $X^{{\rm NS, R}}_r$, $X^{{\rm EW, R}}_r$, $X^{{\rm NS, R}}_r$, and $X^{{\rm EW, R}}_r$ 
fuzzy sets, and $a^{\bm{u},{\rm L}}_{0,r}$, $a^{\bm{u},{\rm L}}_{1,r}$, $a^{\bm{u},{\rm L}}_{2,r}$, 
$a^{\bm{u},{\rm R}}_{0,r}$, $a^{\bm{u},{\rm R}}_{1,r}$, and $a^{\bm{u},{\rm R}}_{2,r}$ 
the tuning parameters.
In the computation of the control inputs, the cumulative states $\bar{\bm{x}}^{{\rm NS, L}}$, 
$\bar{\bm{x}}^{{\rm NS, R}}$, $\bar{\bm{x}}^{{\rm EW, L}}$, and $\bar{\bm{x}}^{{\rm EW, R}}$ 
(i.e., the expected cumulative number of vehicles in the upcoming cycle in the north-south and east-west 
directions of intersections L and R) on the influencing lanes are considered: 
the green times of the northern and southern traffic signals of subnetworks 1 and 2  
are influenced by the number of vehicles on, respectively, the source lanes 2L and 3L, and 2R and 3R 
(see Figure~\ref{fig:network}).

The expected cumulative number of vehicles $\bar{\bm{x}}^{{\rm NS, L}}(k)$ in the upcoming cycle observed on 
lanes 2L and 3L, is the summation of the total number of vehicles 
on these lanes at the current time step and the expected inflow via these source lanes times the cycle time. 
Moreover, $\bar{\bm{x}}^{{\rm EW, L}}(k)$ is the summation of the 
total number of vehicles on lanes 1L and 7L at the current time step and the expected inflow via these source 
lanes times the cycle time. 
The cumulative states  $\bar{\bm{x}}^{{\rm EW, R}}(k)$ and  $\bar{\bm{x}}^{{\rm EW, R}}(k)$ 
are defined in a similar way.%

Next, we discuss the two different control systems that have been designed and evaluated for the 
urban traffic network. 
In order to evaluate and compare the two cases, four different traffic scenarios within 5-min simulations 
were considered. 
Due to the random nature of the simulations (resulting from the inflows of vehicles and the route each vehicle takes 
in the urban traffic network), each scenario was simulated 10 times for each experiment, 
and the average total travel time (TTT) of the vehicles in the urban traffic network was computed.%

\subsubsection{Decentralized fuzzy control}

\begin{figure}
\begin{center}
\includegraphics[width = .415\textwidth]{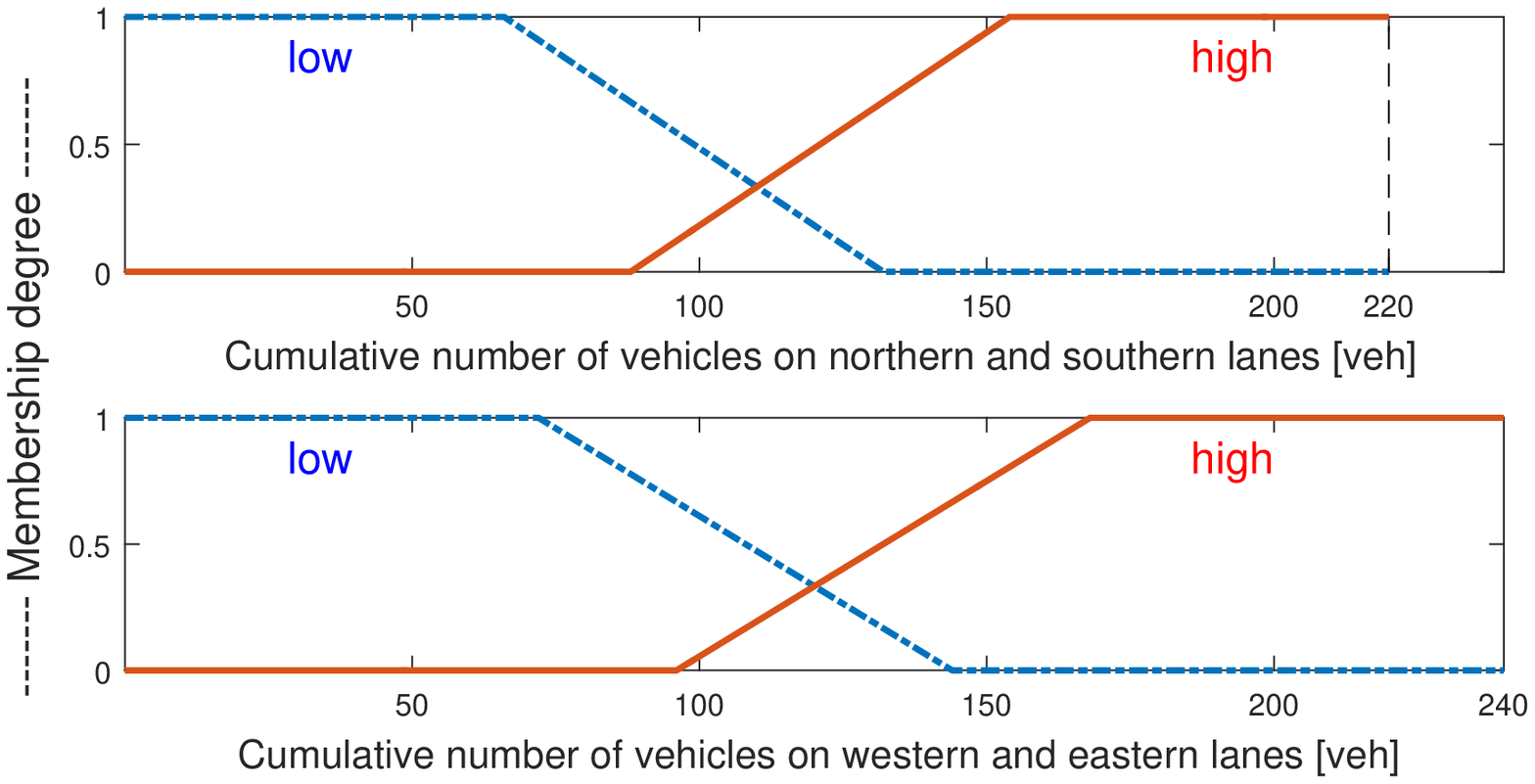}
\caption{Type-1 triangular membership functions for the cumulative
   number of vehicles within the upcoming cycle.
   \label{fig:type_one_MF_control}}
\end{center}
\vspace*{2ex}
%
%
%
\begin{center}
\includegraphics[width = .415\textwidth]{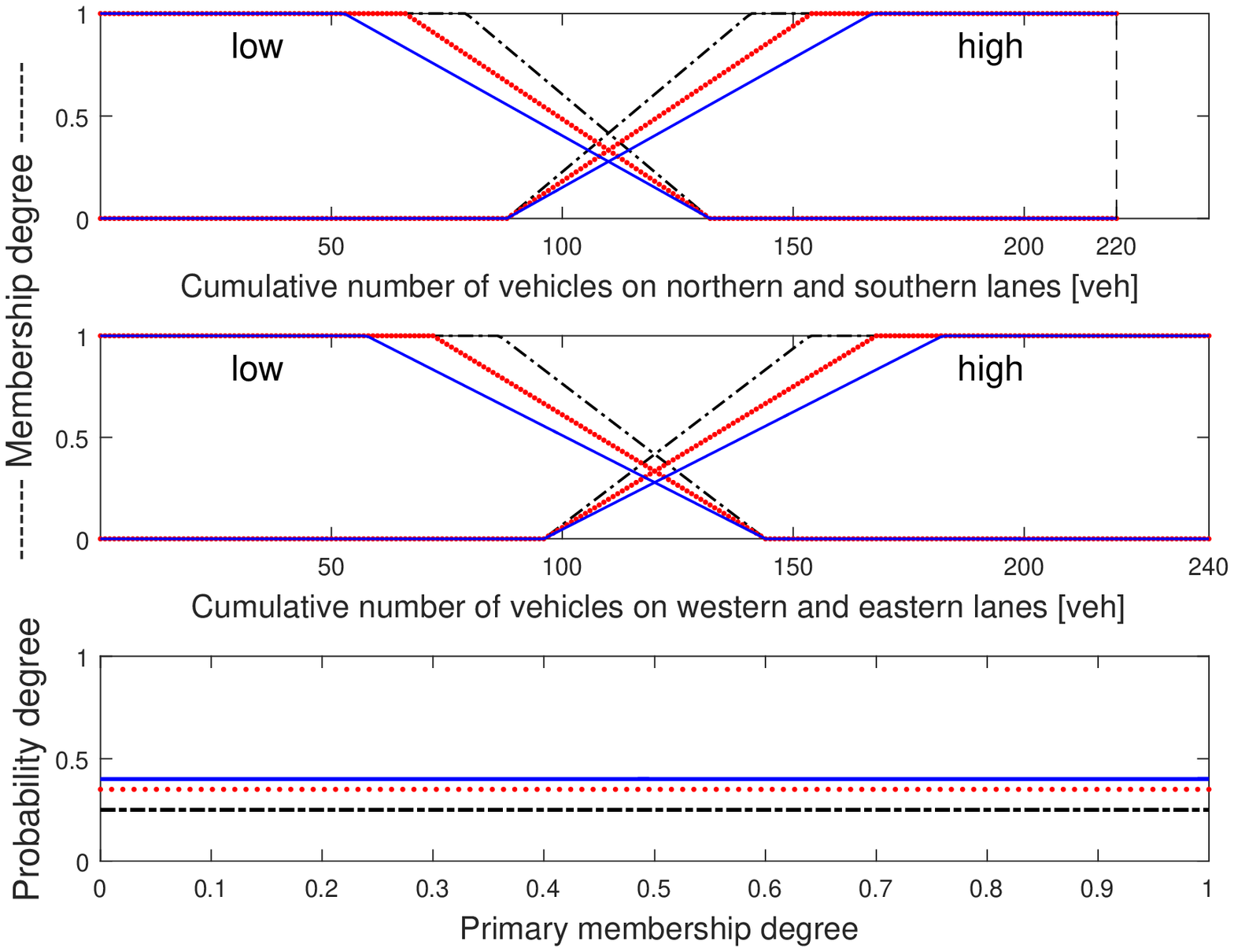}
\caption{Probabilistic-fuzzy membership functions for the cumulative
   number of vehicles within the upcoming cycle.
   \label{fig:Prob_fuzzy_MF_control}}
\end{center}
\end{figure}

First, we consider a decentralized fuzzy control system for the two intersections in 
Figure~\ref{fig:network} (i.e., one fuzzy controller decides about the green time of the 
northern/southern traffic signal of each intersection independently, without coordinating with 
the other fuzzy controller).  
Each fuzzy controller will be tuned individually. 
We will consider two cases: fuzzy controllers that implement type-1 triangular membership functions, 
and fuzzy controllers that use type-2 probabilistic-fuzzy membership functions. 
Note that in modeling the urban traffic network using fuzzy membership functions 
(see Section~\ref{sec:case_study_modeling}), 
type-2 fuzzy-fuzzy membership functions showed to be computationally inefficient or even unfeasible. 
Hence, we have not considered them for control.%

The type-1 triangular and type-2 probabilistic-fuzzy membership functions used for 
the fuzzy controllers are shown in Figures~\ref{fig:type_one_MF_control} and \ref{fig:Prob_fuzzy_MF_control}.  
Table~\ref{table:decentralized_control} shows the results, where the second, third, and fourth columns of this table 
include the TTT for a decentralized  control system  for which each fuzzy controller uses type-1 
triangular membership functions illustrated in Figure~\ref{fig:type_one_MF_control}, 
the TTT for a decentralized  control system for which each fuzzy controller uses type-2 probabilistic-fuzzy  
membership functions illustrated in Figure~\ref{fig:Prob_fuzzy_MF_control}, 
and the relative difference of the TTT.

From Table~\ref{table:decentralized_control}, we see that the decentralized fuzzy control system         
in which each fuzzy controller uses a probabilistic-fuzzy membership function, outperforms the one 
in which the type-1 triangular membership functions are used by the fuzzy controllers.     
In general, the difference between the TTT of the vehicles in the urban traffic network for the 
given four scenarios varies between 100-200~min (which is around 8-15\% of the least TTT 
obtained for each scenario).

\begin{table*}
\begin{center}
\caption{Total travel time (TTT) of the vehicles (given in [min]) for different 5-min simulation 
scenarios with various demand profiles (average over 10 runs for each scenario): 
A comparison between fuzzy controllers with type-1 triangular 
and probabilistic-fuzzy membership functions (MFs).}
\label{table:decentralized_control}
\begin{tabular}{|c|c|c|c|}
\hline 
 Scenario & TTT~[min] for type-1 MF & TTT~[min] for probabilistic-fuzzy MF & relative difference of TTT (\%)\\
\hline
 1 & $1.6 \cdot 10^3$ & $1.4 \cdot 10^3$ & $14.0$\\
 \hline 
2 & $1.7 \cdot 10^3$ & $1.5 \cdot 10^3$ & $13.5$\\
 \hline 
 3 & $1.3 \cdot 10^3$ & $1.2 \cdot 10^3$ & $8.0$\\
 \hline 
4 & $1.8 \cdot 10^3$ & $1.6 \cdot 10^3$ & $12.5$\\
 \hline 
\end{tabular}
\end{center}

%
%
%
\begin{center}
\caption{TTT of the vehicles for different 5-min simulation 
scenarios with various demand profiles (average over 10 runs for each scenario): 
A comparison between decentralized and coordinated fuzzy controllers with 
probabilistic-fuzzy MFs.}
\label{table:coordinated_control}
\begin{tabular}{|c|c|c|c|}
\hline 
 Scenario & TTT~[min] for decentralized control & TTT~[min] for coordinated control & relative difference of TTT (\%)\\
\hline
 1 & $1.4 \cdot 10^3$ & $ 1.3\cdot 10^3$ & $8.0$\\
 \hline 
2 & $1.5 \cdot 10^3$ & $0.8 \cdot 10^3$ & $87.5$\\
 \hline 
 3 & $1.2 \cdot 10^3$ & $1.1 \cdot 10^3$ & $9.0$\\
 \hline 
4 & $1.6 \cdot 10^3$ & $1.4 \cdot 10^3$ & $14.5$\\
 \hline 
\end{tabular}
\end{center}
\end{table*}

\subsubsection{Coordinated predictive-fuzzy control}

Next, we consider a coordinated control system, which includes an MPC module that takes part  
in tuning the fuzzy controllers (see Section~\ref{sec:MPC_controller} for more details). 
Since the MPC module considers the entire traffic network as a whole, under a centralized vision, 
the mutual effects of the dynamics of the two subnetworks on one another will be included in 
the tuned parameters. 
Since the results corresponding to the decentralized control system revealed that the 
probabilistic-fuzzy membership functions outperform the type-1 triangular ones, here only the probabilistic-fuzzy 
membership functions have been considered for the coordinated controllers.%

Table~\ref{table:coordinated_control} illustrates the results corresponding to the coordinated 
control system next to the results obtained for the decentralized control system with probabilistic-fuzzy 
membership functions given in the previous section. 
From these results, we see that compared with the decentralized fuzzy control system, 
the coordinated fuzzy control system gives the least TTT for the vehicles 
in the urban traffic network for all the given four scenarios.  
The difference between the realized values of the TTT for these two control systems 
for the different scenarios varies between 100-700~min (which is around 8-87.5\% of the least TTT obtained for each scenario).%


\section{Conclusions and Future Research}
\label{sec:conclusions}

We have proposed a novel two-layer control architecture that
integrates intelligent and model-predictive control.  The resulting
integrated multi-agent control system is potentially suited for
controlling large-scale and complex-dynamics systems in real time.
This control architecture has a very low computation time, and
provides significant control characteristics, including adaptivity and
coordination, and aims at excellent performance.
Moreover, a general treatment of type-2 fuzzy sets, and
correspondingly type-2 membership functions has been given. This topic
has led to the introduction of two different categories of type-2
fuzzy membership functions (probabilistic-fuzzy, which is a fully
novel concept that has been proposed in the current paper, and
fuzzy-fuzzy), which provide more flexibility and potential in the use
of fuzzy sets and fuzzy membership functions.%

The proposed modeling and control approaches have been implemented to
an urban traffic network.  The results of the case study show that in
modeling the urban traffic network, the probabilistic-fuzzy membership
functions outperform the type-1 triangular and type-2 fuzzy-fuzzy
membership functions, considering both the computation time and
accuracy.  Similarly, probabilistic-fuzzy membership functions provide
the best performance in control.
Moreover, the proposed integrated architecture has a significantly
better performance than a decentralized one that excludes the
coordinating MPC module; more specifically, the total travel time can
be reduced by up to 87.5\% using the proposed control architecture.%



Topics for future research include implementation of a distributed MPC
module (instead of the current centralized setup with only one MPC
controller) in the top control layer, in order to deal with
large-scale systems; and in-depth assessment of stability,
scalability, and attainable performance improvements of the proposed
approach, e.g., via extensive simulations.

\ifCLASSOPTIONcompsoc
\section*{Acknowledgments}
\noindet
\else
\section*{Acknowledgment}
\noindent 
Research supported by a Niels Stensen Fellowship.
\fi

\ifCLASSOPTIONcaptionsoff
  \newpage
\fi

\bibliographystyle{ieeetr}
\bibliography{FuzzyPaper_ref.bib}

\end{document}